\documentclass[aps,prx,reprint,superscriptaddress,longbibliography,nofootinbib,floatfix]{revtex4-2}

\usepackage{amsmath,amssymb,amsthm,bm}
\usepackage{booktabs}
\usepackage{pgfplots}
\pgfplotsset{compat=1.16}
\usepgfplotslibrary{groupplots}
\usetikzlibrary{decorations.markings,arrows.meta}
\usepackage[colorlinks=true,allcolors=blue!55!black]{hyperref}

\theoremstyle{plain}
\newtheorem{theorem}{Theorem}
\newtheorem{proposition}{Proposition}
\newtheorem{lemma}{Lemma}
\newtheorem{corollary}{Corollary}
\theoremstyle{definition}

\newtheorem{example}{Example}
\theoremstyle{remark}
\newtheorem{remark}{Remark}

\newcommand{\kB}{k_{\mathrm{B}}}
\newcommand{\ii}{\mathrm{i}}
\newcommand{\dd}{\mathrm{d}}
\newcommand{\sgn}{\operatorname{sgn}}
\newcommand{\tr}{\operatorname{tr}}
\newcommand{\Lop}{\mathcal{L}}
\newcommand{\Acon}{\mathcal{A}}
\newcommand{\Fcurv}{\mathcal{F}}
\newcommand{\bk}{\bm{k}}
\newcommand{\bv}{\bm{v}}
\newcommand{\bu}{\bm{u}}
\newcommand{\bb}{\bm{b}}
\newcommand{\PT}{\bm{P}_{\!\mathrm{T}}}
\newcommand{\Vc}{V_{\mathrm{cell}}}
\newcommand{\GEP}{G_{\mathrm{EP}}}
\newcommand{\T}{\mathsf{T}}

\begin{document}

\title{Holonomy, Spectral Topology, and Detailed Balance in Classical and Open Quantum Systems}

\author{Alvan Hunt}
\affiliation{Independent researcher, Dublin, Ireland}

\date{\today}

\begin{abstract}
Band topology and non-Hermitian spectral theory are now applied to mechanical metamaterials, chiral active fluids, stochastic networks, and open quantum systems. We give a tutorial synthesis organized by holonomy and by one question: what can the spectral data of a linear generator reveal about detailed balance? Three holonomies recur: the Berry phase of a band, the Burgers vector of a dislocation (the integrated torsion of the defect geometry), and the cycle affinity of a Markov network (the holonomy of the connection $\ln(W_{ij}/W_{ji})$). We rederive the Chern numbers $C_m=m(\sgn\Omega+\sgn\nu_o)$ of a compressible odd-viscous fluid and restate the dislocation zero-mode count of Paulose, Chen, and Vitelli as a pairing of integrated torsion with topological polarization. Our central message is that spectral topology can obstruct detailed balance but does not determine the magnitude of dissipation. Detailed balance is self-adjointness in the inner product fixed by the stationary state, so a non-real eigenvalue, an exceptional point (EP), or a nonzero point-gap winding number of a Markov generator rules it out and forces entropy production $\sigma>0$, with $\sigma\ge\kB j^2/D$ for any current, whereas a real, diagonalizable spectrum does not in general certify it. In the Hatano-Nelson ring, which is normal yet has complex spectrum, $\sigma$ sweeps $(0,\infty)$ at fixed winding number. EPs need a separate assessment of the dynamics: with generalized detailed balance including time-reversal parity, an EP is compatible with equilibrium only through time-reversal-odd dynamics, as at critical damping of a Langevin oscillator, whereas an EP of a system with only even variables, such as an odd elastic medium, certifies a nonequilibrium steady state; in both cases $\sigma$ is analytic across the EP. We propose benchtop tests for each statement.
\end{abstract}

\maketitle

\section{Introduction}
\label{sec:intro}

The recognition that band structures can carry integer invariants~\cite{TKNN1982,Simon1983,Berry1984,HasanKane2010} has, over the past fifteen years, been exported far beyond electrons. Topological phonons were identified in microtubules~\cite{Prodan2009} and isostatic lattices~\cite{Kane2014}, then built in gyroscopic and pendulum metamaterials~\cite{Nash2015,Susstrunk2015,Huber2016,Mao2018}. Chiral active fluids, in which odd viscosity~\cite{Avron1995,Avron1998,Banerjee2017} breaks time-reversal symmetry without dissipation, support topologically protected sound~\cite{Souslov2017,Shankar2017,Souslov2019,Shankar2022}. Stochastic networks driven out of equilibrium carry chiral edge currents in configuration space~\cite{Murugan2017,Dasbiswas2018,Tang2021}. In parallel, the spectral theory of non-Hermitian operators, with its exceptional points (EPs)~\cite{Kato1966,Heiss2012,Miri2019}, its point-gap invariants, and its skin effect~\cite{Hatano1996,Yao2018,Gong2018,Kawabata2019,Okuma2020,Zhang2020}, has become a field in its own right~\cite{Ashida2020,Bergholtz2021,Coulais2021}.

These literatures share words (\emph{topological}, \emph{chiral}, \emph{non-Hermitian}, \emph{torsion}) whose meanings overlap only partially. That overlap invites two opposite errors: treating a shared formalism as evidence of shared physics, and missing a genuinely common structure because each community phrases it differently. This article is a perspective and tutorial aimed at both. Its organizing device is holonomy, and its central message concerns detailed balance.

\emph{Holonomy.} In each system an observable is the result of transporting something around a closed loop: a Bloch state around a cycle of the Brillouin zone (the Berry phase)~\cite{Simon1983,Berry1984}, a lattice frame around a dislocation core (the Burgers vector)~\cite{Bilby1955,Katanaev1992,Katanaev2005}, or a probability flux around a cycle of a Markov network (the cycle affinity)~\cite{Schnakenberg1976}. In each case the holonomy is gauge invariant, can be written as the flux of a curvature or torsion through a surface bounded by the loop, and is tied by an index theorem or a conservation law to something observable at a boundary, a defect, or in a current.

\emph{Detailed balance as self-adjointness in a physical inner product.} Each system has a linear generator $\Lop$ of its dynamics: a dynamical matrix, a linearized hydrodynamic operator, a rate matrix, a Langevin drift matrix, or a Lindbladian. The physics singles out an inner product, set by the energy for conservative dynamics and by the stationary state for stochastic dynamics. For stochastic dynamics whose variables are all even under time reversal, detailed balance is precisely self-adjointness of $\Lop$ in the stationary inner product. Spectral data (non-real eigenvalues, EPs, spectral winding) can show that $\Lop$ is self-adjoint in no inner product at all, and therefore obstruct detailed balance. In general they cannot certify it, because self-adjointness in \emph{some} inner product, which is all a real, diagonalizable spectrum provides in general, is weaker than self-adjointness in the stationary one; special spectra are an exception (Sec.~\ref{sec:thermo}). Nor is normality, the commutation of $\Lop$ with its adjoint, the relevant property: a normal generator can be far from equilibrium. When some variables are odd under time reversal, as momenta are, equilibrium permits an anti-self-adjoint part of $\Lop$, and EPs become compatible with detailed balance.

The resulting message is short: \emph{spectral topology can obstruct detailed balance, but it does not determine the magnitude of dissipation, and exceptional points require a separate assessment of the underlying dynamics.} Table~\ref{tab:dictionary} collects the holonomies. Our specific contributions are: (a) a tutorial rederivation of the Chern numbers of a compressible odd-viscous fluid, Eq.~\eqref{eq:chern}, reproducing Refs.~\cite{Souslov2019,Tauber2019}, including the case without odd viscosity, where the integrated curvature is an integer but not an invariant; (b) a restatement of the dislocation zero-mode count of Ref.~\cite{Paulose2015} as a pairing between integrated torsion and the topological polarization, Eq.~\eqref{eq:torsion-count}; (c) a topology-dissipation theorem with sharply stated scope, Theorem~\ref{thm:topodiss}, together with an explicit example showing that its converse fails, Example~\ref{ex:3state}, and an exact demonstration, using a ring that saturates a known affinity-dependent bound~\cite{Pietzonka2016b}, that entropy production is not fixed by an integer invariant, Eq.~\eqref{eq:gcothg}, placed next to known quantitative spectral bounds~\cite{Barato2017,Kolchinsky2024}; and (d) a classification of EPs relative to generalized detailed balance, Proposition~\ref{prop:GDB}, anchored on the critically damped Langevin oscillator, with a closed-form entropy production for an odd elastic mode that is analytic across its EP, Eq.~\eqref{eq:oddsigma}. Items (a) and (b) are established results presented in a common language; the contribution is the synthesis and the precise statement of what does and does not transfer between systems. Symbolic results and counterexamples were checked in Wolfram Language (Appendix~\ref{app:verification}).

We make no claim that the phenomena discussed here are manifestations of a single microscopic field or of one geometry seen at different energy scales. Shared mathematics is not shared physics; where an analogy is only formal (for example, between crystal torsion and the torsion of Einstein-Cartan gravity~\cite{Hehl1976}) we say so.

Section~\ref{sec:architecture} sets up generators, inner products, and bundles, and works through the odd-viscous fluid. Section~\ref{sec:defects} treats dislocations as torsion. Section~\ref{sec:thermo} introduces the thermodynamic connection and proves the topology-dissipation theorem. Section~\ref{sec:EP} analyzes EPs relative to generalized detailed balance. Section~\ref{sec:experiments} proposes tests, and Sec.~\ref{sec:discussion} discusses scope and open problems.

\begin{table*}[t]
\caption{\label{tab:dictionary}Dictionary of the geometric structures used in this paper. Each row is a connection (or, in the last row, a spectral structure) whose loop invariant is observable at a boundary, a defect, or in a current.}
\begin{ruledtabular}
\footnotesize
\begin{tabular}{p{2.15cm}p{2.75cm}p{3.3cm}p{3.4cm}p{3.2cm}c}
\raggedright Structure &\raggedright Base space and loops &\raggedright Connection &\raggedright Holonomy (gauge invariant) &\raggedright Signature & Sec.\tabularnewline
\colrule
\raggedright Bloch bundle &\raggedright Brillouin zone $T^2$ (or compactified $\mathbb{R}^2$) &\raggedright $\Acon_n=\ii\langle u_n|\dd u_n\rangle$ &\raggedright Berry phase; Chern number $C_n$ &\raggedright chiral edge modes & \ref{sec:architecture}\tabularnewline
\raggedright Defect geometry &\raggedright material plane; loops around cores &\raggedright co-frame $e^a$, spin connection $\omega^a{}_b$ &\raggedright Burgers vector $b^a=\oint e^a=\int T^a$ &\raggedright dislocation-bound zero modes & \ref{sec:defects}\tabularnewline
\raggedright Thermodynamic &\raggedright state graph; cycles $\gamma$ &\raggedright $\Acon_{ij}=\ln(W_{ij}/W_{ji})$ &\raggedright cycle affinity $\Acon[\gamma]$ &\raggedright currents and $\sigma>0$; skin accumulation & \ref{sec:thermo}\tabularnewline
\raggedright Spectral &\raggedright parameter space of $\Lop$ &\raggedright (none) &\raggedright eigenvalue exchange on encircling an EP &\raggedright thresholds; chiral state transfer & \ref{sec:EP}\tabularnewline
\end{tabular}
\end{ruledtabular}
\end{table*}

\section{Generators, inner products, and bundles}
\label{sec:architecture}

\subsection{Linear generators}

At the level of linear response about a steady state, every system considered here obeys
\begin{equation}
\partial_t x = \Lop x
\label{eq:generic}
\end{equation}
on a finite-dimensional space, or on one Fourier sector of a translation-invariant medium, possibly with additive noise. Five generators recur.

\emph{(a) Conservative and overdamped mechanics.} Displacements $\bu$ of a lattice with mass matrix $M$ obey $M\ddot{\bu}=-D\bu$ with $D=QKQ^{\T}$, where $Q$ is the equilibrium matrix and $K$ the diagonal matrix of spring constants~\cite{Kane2014,Mao2018}. With $x=(M^{1/2}\dot{\bu},K^{1/2}Q^{\T}\bu)$ one finds $\Lop=\left(\begin{smallmatrix}0&-B\\ B^{\T}&0\end{smallmatrix}\right)$, $B=M^{-1/2}QK^{1/2}$, a real antisymmetric matrix, and $|x|^2$ is twice the energy; $\ii\Lop$ is Hermitian and is the ``square root'' of $D$ exploited in topological mechanics. In the overdamped limit, $\zeta\dot{\bu}=-D\bu$ with a positive friction matrix $\zeta$.

\emph{(b) Linearized hydrodynamics.} Fields $\psi(\bm r,t)$ obey $\ii\partial_t\psi=H(-\ii\nabla)\psi$, so that $\Lop=-\ii H(\bk)$ in each Fourier sector.

\emph{(c) Markov jump processes.} Probabilities $p_i$ on a finite state graph obey $\dot p=Wp$, with $W_{ij}\ge0$ the rate of $j\to i$ ($i\neq j$) and $W_{jj}=-\sum_{i\neq j}W_{ij}$.

\emph{(d) Lindblad dynamics.} Density matrices obey $\dot\rho=\Lop\rho=-\ii[H,\rho]+\sum_k\big(L_k\rho L_k^\dagger-\tfrac12\{L_k^\dagger L_k,\rho\}\big)$.

\emph{(e) Linear Langevin dynamics.} Real variables obey $\dd x=Ax\,\dd t+\sqrt2\,B\,\dd W$ with noise matrix $\Xi=BB^{\T}\ge0$. The mean obeys $\partial_t\langle x\rangle=A\langle x\rangle$, and for Hurwitz $A$ (all eigenvalues with negative real part) the stationary covariance $\Sigma$ solves $A\Sigma+\Sigma A^{\T}+2\Xi=0$.

The most useful elementary fact about such generators is the following.

\begin{lemma}[Self-adjointness excludes EPs]
\label{lem:selfadjoint}
For a linear operator $\Lop$ on $\mathbb{C}^n$ the following are equivalent: (i) $\Lop$ is diagonalizable with real spectrum; (ii) there is a positive-definite $G$ with $G\Lop=\Lop^\dagger G$, i.e., $\Lop$ is self-adjoint in the inner product $\langle x,y\rangle_G=x^\dagger Gy$. Likewise, $\Lop$ is diagonalizable with purely imaginary spectrum if and only if $G\Lop=-\Lop^\dagger G$ for some $G>0$.
\end{lemma}

\begin{proof}
If (ii) holds, $G^{1/2}\Lop G^{-1/2}$ is Hermitian. If (i) holds with $\Lop=V\Lambda V^{-1}$, take $G=(VV^\dagger)^{-1}$. The anti-self-adjoint case is identical with $\Lambda$ imaginary.
\end{proof}

Consequently, a generator with an EP or a non-real eigenvalue is self-adjoint in no inner product. Two distinctions are used throughout.

\emph{Self-adjointness is not normality.} $\Lop$ is normal in a given inner product if it commutes with its adjoint. Normal operators are unitarily diagonalizable and so have no EPs, but their eigenvalues may be complex: the Hatano-Nelson ring of Sec.~\ref{sec:HN} is normal, has complex spectrum and nonzero spectral winding, and produces entropy. Normality is therefore not the relevant diagnostic: self-adjointness in the physical inner product implies normality there, but normality alone is compatible with dissipation.

\emph{The physical inner product is not just any inner product.} Physical constraints refer to one inner product. For conservative mechanics in the variables of (a), it is the energy norm, and $\Lop$ is anti-self-adjoint there. For stochastic dynamics it is fixed by the stationary state, and detailed balance is self-adjointness in that inner product (Proposition~\ref{prop:DB} and Sec.~\ref{sec:GDB}). Lemma~\ref{lem:selfadjoint} shows that a real, diagonalizable spectrum guarantees self-adjointness only in \emph{some} inner product, which need not be the physical one (Example~\ref{ex:3state}).

\subsection{Bundles, connections, and invariants}

For a translation-invariant system, $\Lop(\bk)$ depends smoothly on crystal momentum $\bk$ in the Brillouin zone $T^d$. For Hermitian $H(\bk)$ and an isolated band $n$ with eigenprojector $P_n(\bk)$, the image of $P_n$ defines a line bundle over $T^d$ with Berry connection $\Acon_n=\ii\langle u_n|\dd u_n\rangle$ and curvature $\Fcurv_n=\dd\Acon_n$~\cite{Simon1983,Berry1984}. The holonomy around a closed loop $\ell$ is the Berry phase $\gamma_n[\ell]=\oint_\ell\Acon_n$, and in $d=2$ the total curvature is the first Chern number
\begin{equation}
\begin{gathered}
C_n=\frac{1}{2\pi}\int_{T^2}\Fcurv_n,\\
\Fcurv_{n,xy}=-2\,\mathrm{Im}\sum_{m\neq n}\frac{\langle n|\partial_{k_x}H|m\rangle\langle m|\partial_{k_y}H|n\rangle}{(E_n-E_m)^2},
\end{gathered}
\label{eq:berry}
\end{equation}
where the second form is manifestly gauge invariant~\cite{TKNN1982,HasanKane2010}. Since the band projector is a spectral projection of $H(\bk)$, we use ``spectral topology'' for both this eigenvector topology and the point-gap topology of the spectrum itself.

For non-Hermitian $\Lop(\bk)$ two notions of gap are distinguished~\cite{Gong2018,Kawabata2019}. If the spectrum avoids a line in the complex plane (line gap), $\Lop(\bk)$ can be continuously deformed to a Hermitian or anti-Hermitian operator without closing the gap, and its invariants are those of band theory, computed with biorthogonal projectors. If the spectrum avoids a point $E$ (point gap), there is an intrinsically non-Hermitian invariant, which in $d=1$ is the spectral winding number
\begin{equation}
w(E)=\frac{1}{2\pi\ii}\oint_0^{2\pi}\dd k\,\partial_k\ln\det\big[\Lop(k)-E\big].
\label{eq:winding}
\end{equation}
A nonzero $w(E)$ under periodic boundary conditions implies the skin effect, an extensive accumulation of eigenmodes at one boundary under open boundary conditions~\cite{Yao2018,Okuma2020,Zhang2020}. A nonzero $w$ requires $\Lop(k)$ to have non-real eigenvalues for some $k$, since a closed curve of real eigenvalues winds around no point; it does not require non-normality.

In continuous media $\bk\in\mathbb{R}^2$ is not compact. The integrated curvature is a topological invariant when the band projectors $P_n(\bk)$ approach a direction-independent limit as $|\bk|\to\infty$, so that the bundle extends to the one-point compactification $S^2$. The operator $H(\bk)$ itself need not converge; in the example below it grows quadratically. Odd viscosity supplies exactly this regularization~\cite{Souslov2019,Tauber2019}.

\subsection{Worked example: the compressible odd-viscous fluid}
\label{sec:oddfluid}

Consider a two-dimensional compressible fluid with sound speed $c$, linearized about density $\rho_0$ at rest:
\begin{align}
\partial_t\delta\rho &= -\rho_0\nabla\cdot\bv,\nonumber\\
\partial_t\bv &= -\frac{c^2}{\rho_0}\nabla\delta\rho + \Omega\,\bv^* + \nu_o\nabla^2\bv^*,
\label{eq:oddfluid}
\end{align}
with $\bv^*\equiv\epsilon\cdot\bv=(v_y,-v_x)$, $\epsilon_{xy}=-\epsilon_{yx}=1$. The term $\Omega\bv^*$ is a Coriolis-type force: in a frame rotating counterclockwise at $\Omega_r$, the Coriolis acceleration $-2\Omega_r\hat{\bm z}\times\bv$ equals $2\Omega_r\bv^*$, so $\Omega=2\Omega_r$. The term $\nu_o\nabla^2\bv^*$, with $\nu_o=\eta_o/\rho_0$, is the divergence of the odd stress $\sigma^o_{ij}=\eta_o(\partial_iv^*_j+\partial^*_iv_j)$, $\partial^*_i=\epsilon_{ik}\partial_k$~\cite{Avron1998,Banerjee2017}. Because the odd viscosity tensor is antisymmetric under exchange of index pairs, $\eta^o_{ijkl}=-\eta^o_{klij}$, it does no work: its contribution to the dissipation rate $\eta_{ijkl}\partial_iv_j\partial_kv_l$ vanishes identically. Two remarks are useful.

\begin{remark}[Incompressible limit]
For $\nabla\cdot\bv=0$ one has, identically, $\nabla^2\bv^*=\nabla\omega$ with vorticity $\omega=\partial_xv_y-\partial_yv_x$. The odd force is then a gradient and is absorbed into the modified pressure $p-\eta_o\omega$~\cite{Avron1998,Ganeshan2017}. Odd viscosity is invisible in bulk incompressible flow and acts through compressibility or boundaries, as observed in spinning colloidal fluids~\cite{Soni2019}.
\end{remark}

\begin{remark}[Hermiticity]
In Fourier space with $\psi=(c\,\delta\rho/\rho_0,v_x,v_y)$ and $\psi\propto e^{\ii(\bk\cdot\bm r-\omega t)}$, Eqs.~\eqref{eq:oddfluid} become $\omega\psi=H(\bk)\psi$ with
\begin{equation}
H(\bk)=\begin{pmatrix}0&ck_x&ck_y\\ ck_x&0&\ii f(k)\\ ck_y&-\ii f(k)&0\end{pmatrix},\qquad f(k)=\Omega-\nu_o k^2.
\label{eq:Hodd}
\end{equation}
$H(\bk)$ is Hermitian in the energy norm, as it must be for non-dissipative dynamics.
\end{remark}

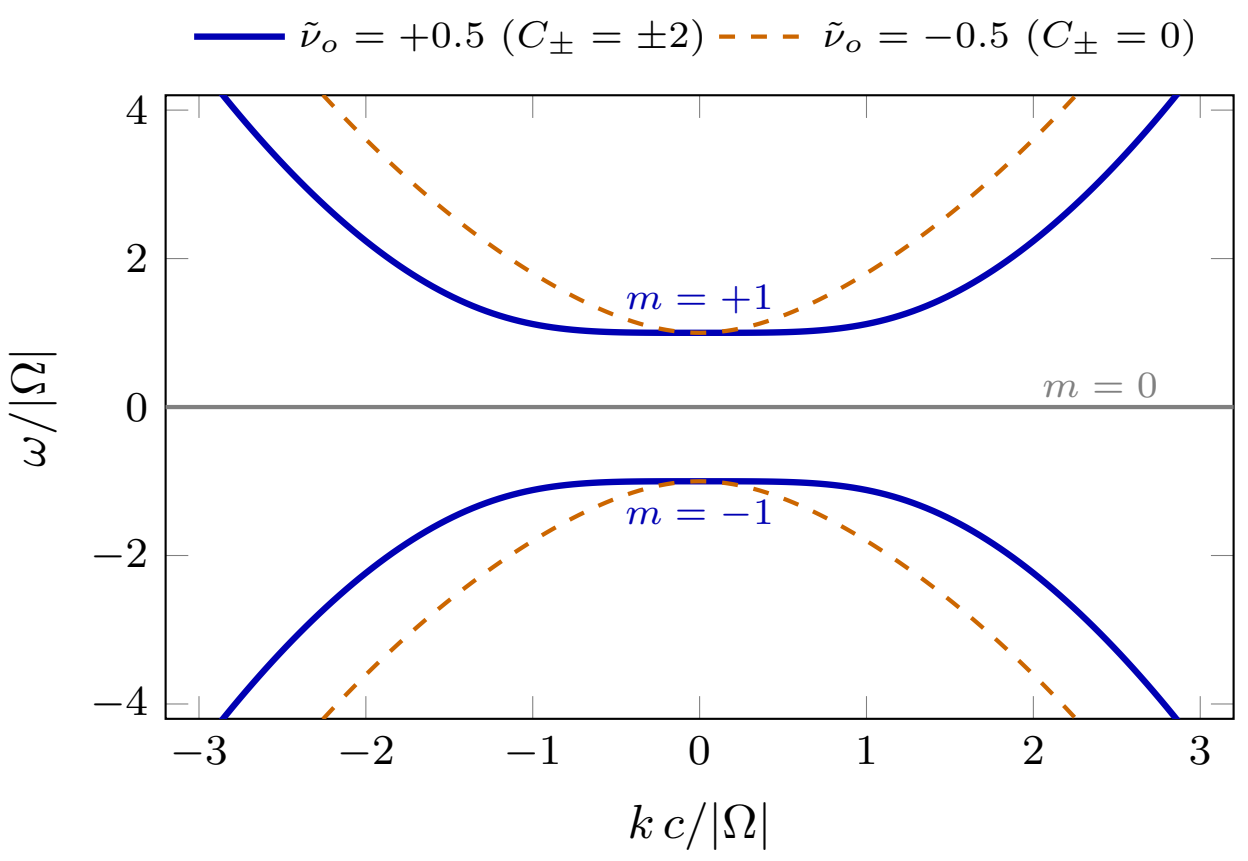
\begin{figure}[t]
\centering
\begin{tikzpicture}
\begin{axis}[
  width=\columnwidth, height=0.66\columnwidth,
  xlabel={$k\,c/|\Omega|$}, ylabel={$\omega/|\Omega|$},
  xmin=-3.2, xmax=3.2, ymin=-4.2, ymax=4.2,
  domain=-3.2:3.2, samples=161,
  tick label style={font=\footnotesize}, label style={font=\small},
  legend style={font=\scriptsize, at={(0.5,1.03)}, anchor=south, draw=none, fill=none, legend columns=2},
  legend cell align=left]
\addplot[very thick, blue!70!black] {sqrt(x^2+(1-0.5*x^2)^2)};
\addlegendentry{$\tilde\nu_o=+0.5$ ($C_\pm=\pm2$)}
\addplot[thick, dashed, orange!80!black] {sqrt(x^2+(1+0.5*x^2)^2)};
\addlegendentry{$\tilde\nu_o=-0.5$ ($C_\pm=0$)}
\addplot[very thick, blue!70!black, forget plot] {-sqrt(x^2+(1-0.5*x^2)^2)};
\addplot[thick, dashed, orange!80!black, forget plot] {-sqrt(x^2+(1+0.5*x^2)^2)};
\addplot[thick, gray, forget plot] {0};
\node[font=\scriptsize, blue!70!black] at (axis cs:0,1.45) {$m=+1$};
\node[font=\scriptsize, gray] at (axis cs:2.4,0.3) {$m=0$};
\node[font=\scriptsize, blue!70!black] at (axis cs:0,-1.45) {$m=-1$};
\end{axis}
\end{tikzpicture}
\caption{\label{fig:bands}Bands of the odd-viscous fluid, Eq.~\eqref{eq:Hodd}, for $\Omega>0$ and dimensionless odd viscosity $\tilde\nu_o=\nu_o\Omega/c^2=\pm0.5$. The spectra differ only quantitatively, yet the bands are topological ($C_\pm=\pm2$) only when $\sgn\Omega=\sgn\nu_o$ (solid), Eq.~\eqref{eq:chern}. The invariant is a property of the eigenvector bundle, not of the dispersion.}
\end{figure}

The spectrum is $\omega_0=0$ (the geostrophic, vortical band) and $\omega_\pm=\pm\sqrt{c^2k^2+f(k)^2}$, gapped for $\Omega\neq0$ (Fig.~\ref{fig:bands}). Writing $H=ck_x\Lambda_1+ck_y\Lambda_2+f\Lambda_3$, the three matrices obey $[\Lambda_a,\Lambda_b]=-\ii\epsilon_{abc}\Lambda_c$ and $\sum_a\Lambda_a^2=2\cdot\mathbb{1}$, so $-\Lambda_a$ form a spin-1 representation (Appendix~\ref{app:spin1}). Band $m\in\{-1,0,+1\}$ (labelled by the sign of its frequency) therefore carries Berry curvature equal, up to an orientation sign, to $m$ times the solid-angle density swept by the unit vector $\hat{\bm d}=(ck_x,ck_y,f)/|\bm d|$, so that each wrapping of the sphere ($4\pi$ of solid angle) contributes $2m$ to the Chern number. The projectors depend on $\bk$ only through $\hat{\bm d}$. At $k=0$, $\hat{\bm d}$ points along $\sgn(\Omega)\hat{\bm z}$; as $k\to\infty$ it points along $-\sgn(\nu_o)\hat{\bm z}$ in every direction, so the bundle extends to $S^2$. The map $\hat{\bm d}:S^2\to S^2$ has degree $\pm1$ if $\sgn\Omega=\sgn\nu_o$ and degree zero otherwise. Evaluating Eq.~\eqref{eq:berry} gives
\begin{equation}
C_m = m\,\big(\sgn\Omega+\sgn\nu_o\big),\qquad m\in\{-1,0,+1\},
\label{eq:chern}
\end{equation}
which we confirmed by direct numerical integration of Eq.~\eqref{eq:berry} for all four sign combinations (Appendix~\ref{app:spin1}). The bands are topological, with $C_\pm=\pm2$, only when the Coriolis frequency and the odd viscosity have the same sign; this reproduces Refs.~\cite{Souslov2019,Tauber2019}.

Two cautions belong in any tutorial treatment. First, without odd viscosity ($\nu_o=0$), $\hat{\bm d}\to(\cos\theta,\sin\theta,0)$ as $k\to\infty$, a direction-dependent limit. The curvature then integrates over a hemisphere of directions and gives $\pm\sgn\Omega$ for the $\pm$ bands, an integer by geometric coincidence rather than a topological invariant: an arbitrarily small odd viscosity changes it to $\pm2\sgn\Omega$ or $0$, depending on the sign of $\nu_o$ (Appendix~\ref{app:spin1}). Second, in continuous media the bulk-edge correspondence can fail: the number of edge branches crossing a gap depends on the boundary condition and need not equal the difference of bulk Chern numbers~\cite{Tauber2020,Graf2021}. Equation~\eqref{eq:chern} is a statement about the bulk bundle; the edge spectrum must be computed for the boundary in question.

\section{Defect geometry: dislocations as torsion}
\label{sec:defects}

\subsection{Cartan structure equations of a crystal}

The geometric theory of defects describes a crystal by a co-frame $e^a=e^a{}_\mu\dd x^\mu$, which maps material directions to local lattice directions, and a spin connection $\omega^a{}_b$, which records lattice rotations~\cite{Bilby1955,Katanaev1992,Katanaev2005}. Torsion and curvature are defined by the Cartan structure equations,
\begin{equation}
T^a=\dd e^a+\omega^a{}_b\wedge e^b,\qquad R^a{}_b=\dd\omega^a{}_b+\omega^a{}_c\wedge\omega^c{}_b .
\label{eq:cartan}
\end{equation}
The identification, due to Kondo, Bilby, Kr\"oner and others and systematized by Katanaev and Volovich~\cite{Bilby1955,Katanaev1992}, is that the torsion 2-form is the dislocation density and the curvature 2-form is the disclination density. These are the same structure equations as in Einstein-Cartan gravity, but the analogy is formal: in gravity, torsion is sourced algebraically by spin density and does not propagate~\cite{Hehl1976}, whereas in a crystal it is the record of plastic history.

\subsection{The Burgers vector as a holonomy}

In a defect-free crystal $e^a=\dd y^a$, where $y^a(\bm x)=x^a+u^a(\bm x)$ are lattice-adapted coordinates, and $\omega=0$; then $T^a=\dd^2y^a=0$. With dislocations, $e^a$ remains smooth and single valued away from the cores but is no longer exact. In a disclination-free medium we may choose the teleparallel gauge $\omega^a{}_b=0$, and Stokes' theorem gives, for any surface $\mathcal{S}$ with boundary $\partial\mathcal{S}$,
\begin{equation}
b^a\equiv\oint_{\partial\mathcal{S}}e^a=\int_{\mathcal{S}}\dd e^a=\int_{\mathcal{S}} T^a .
\label{eq:burgers}
\end{equation}
The Burgers vector is thus the translational holonomy of the frame, and an isolated dislocation at $\bm x_0$ is a delta-function torsion flux, $T^a=b^a\,\delta^2(\bm x-\bm x_0)\,\dd x\wedge\dd y$. (With disclinations present, translational and rotational holonomies mix and $b^a$ depends on a base point; we exclude this case. For a bias locked to the frame, the rotational holonomy also means that the bias cannot be single-valued around the core and must jump across a seam; the consequences for dissipation will be discussed elsewhere.) Equation~\eqref{eq:burgers} is an identity, and it is the precise sense in which ``torsion'' enters mechanical metamaterials.

\subsection{Pairing torsion with band topology}

Maxwell lattices, with equal numbers of degrees of freedom and constraints, carry a topological invariant built from the compatibility matrix $Q(\bk)$~\cite{Kane2014,Mao2018}. The winding numbers $n_i=(2\pi\ii)^{-1}\oint_{\mathcal C_i}\dd\ln\det Q(\bk)$ along the two non-contractible cycles of the Brillouin zone define the topological polarization $\PT=\sum_in_i\bm a_i$, a lattice vector. Paulose, Chen and Vitelli showed that the topological contribution to the number of zero modes minus states of self-stress localized at a dislocation is $\nu_T=\PT\cdot\bm d/\Vc$, where $\bm d$ is the Burgers vector rotated by $\pi/2$~\cite{Paulose2015}; a local contribution, set by coordination near the core, adds to it. Combining their result with Eq.~\eqref{eq:burgers} restates it in the language of defect geometry:
\begin{equation}
\nu_T=\frac{1}{\Vc}\,\PT\cdot R_{\pi/2}\int_{\mathcal{S}}\bm T ,
\label{eq:torsion-count}
\end{equation}
where $\mathcal{S}$ encloses the core and $R_{\pi/2}$ is the rotation. Because $\PT$ and $\bb$ are both lattice vectors, $\PT\cdot R_{\pi/2}\bb=\pm\hat{\bm z}\cdot(\bb\times\PT)$ is an integer multiple of $\Vc$, so $\nu_T\in\mathbb{Z}$.

Equation~\eqref{eq:torsion-count} contains no physics beyond Ref.~\cite{Paulose2015}. Its value here is structural: a holonomy of the defect geometry (integrated torsion) paired with a holonomy of the band bundle (the winding numbers in $\PT$) yields a protected, integer count of localized modes. An analogous pairing between the Burgers vector and the weak indices governs helical modes on dislocations in three-dimensional topological insulators~\cite{Ran2009,Teo2010}, and dislocations bind non-Hermitian skin modes when the bulk carries weak point-gap invariants~\cite{Schindler2021,Bhargava2021}. A dislocation dipole carries opposite counts at its two cores, and reversing $\PT$, for instance by a Guest-Hutchinson soft deformation of a transformable kagome lattice~\cite{Rocklin2017}, exchanges zero modes and states of self-stress between them. Section~\ref{sec:experiments} turns this into a test.

\section{The thermodynamic connection and a topology-dissipation theorem}
\label{sec:thermo}

\subsection{Setting}

Let $W$ generate an irreducible Markov jump process on a finite connected graph whose states are even under time reversal, with $W_{ij}>0$ if and only if $W_{ji}>0$ (microscopic reversibility) and unique stationary state $\pi>0$. Stationary edge currents are $J_{ij}=W_{ij}\pi_j-W_{ji}\pi_i$, and the entropy production rate is~\cite{Schnakenberg1976,Seifert2012}
\begin{equation}
\sigma=\frac{\kB}{2}\sum_{i\neq j}J_{ij}\ln\frac{W_{ij}\pi_j}{W_{ji}\pi_i}\ \ge 0 .
\label{eq:sigma}
\end{equation}

\subsection{The rate connection}

On each oriented edge define
\begin{equation}
\Acon_{ij}=\ln\frac{W_{ij}}{W_{ji}}=-\Acon_{ji}.
\label{eq:rateconnection}
\end{equation}
This is a discrete $\mathbb{R}$-valued connection (a 1-cochain) on the state graph, introduced in this sense by Polettini~\cite{Polettini2012}, who identified its gauge freedom and showed that its Wilson loops are Schnakenberg's affinities. A diagonal similarity transformation $W\mapsto\Phi^{-1}W\Phi$, $\Phi=\mathrm{diag}(\phi_i)>0$, leaves the spectrum invariant and shifts $\Acon_{ij}\mapsto\Acon_{ij}+2(\ln\phi_j-\ln\phi_i)$, an exact cochain; these are its gauge transformations. The transformed matrix generally no longer has zero column sums, so it represents the same dynamics in other coordinates rather than another Markov generator in the same probability coordinates (for $\Phi=\Pi^{1/2}$ and detailed balance it is the symmetrized form); what every such representation shares is the holonomy. The gauge-invariant content is the holonomy around cycles $\gamma=(i_0\to i_1\to\cdots\to i_0)$,
\begin{equation}
\Acon[\gamma]=\sum_{k}\Acon_{i_{k+1}i_k}=\ln\prod_k\frac{W_{i_{k+1}i_k}}{W_{i_ki_{k+1}}},
\end{equation}
which is Schnakenberg's cycle affinity~\cite{Schnakenberg1976}. Since $\ln(W_{ij}\pi_j/W_{ji}\pi_i)=\Acon_{ij}+\ln\pi_j-\ln\pi_i$ and a divergence-free stationary current has zero pairing with any exact cochain,
\begin{equation}
\sigma=\kB\sum_{\text{edges }e}J_e\,\Acon_e=\kB\sum_{\gamma\in\text{basis}}J_\gamma\,\Acon[\gamma],
\label{eq:sigmapairing}
\end{equation}
where the second sum runs over a basis of fundamental cycles with cycle currents $J_\gamma$. Dissipation is the pairing of the current with the holonomy of the rate connection.

\begin{proposition}[Detailed balance as flatness]
\label{prop:DB}
The following are equivalent: (i) detailed balance, $W_{ij}\pi_j=W_{ji}\pi_i$ for all edges; (ii) $\Acon$ is exact, $\Acon_{ij}=\varphi_i-\varphi_j$ for some $\varphi$; (iii) every cycle affinity vanishes (Kolmogorov's criterion~\cite{Kolmogorov1936}); (iv) $\sigma=0$; (v) $W$ is self-adjoint in the stationary inner product $\langle x,y\rangle_\pi=\sum_ix_iy_i/\pi_i$.
\end{proposition}

\begin{proof}
(i)$\Rightarrow$(ii) with $\varphi=\ln\pi$. (ii)$\Rightarrow$(iii) since exact cochains have zero circulation. (iii)$\Rightarrow$(ii) by path integration from a root vertex, which is path independent. (ii)$\Rightarrow$(i): $\tilde\pi_i\propto e^{\varphi_i}$ satisfies detailed balance, hence is stationary, hence equals $\pi$ by uniqueness. (i)$\Leftrightarrow$(iv): each pair of terms in Eq.~\eqref{eq:sigma} has the form $(x-y)\ln(x/y)\ge0$, with equality if and only if $x=y$. (i)$\Leftrightarrow$(v): $\langle x,Wy\rangle_\pi=\langle Wx,y\rangle_\pi$ for all $x,y$ if and only if $W_{ij}/\pi_i=W_{ji}/\pi_j$.
\end{proof}

Item (v) refers to one inner product. Self-adjointness in some other inner product is strictly weaker, as the following example shows.

\begin{example}[A real, diagonalizable spectrum does not certify detailed balance]
\label{ex:3state}
Take
\begin{equation}
W=\begin{pmatrix}-3&1&1\\2&-4&1\\1&3&-2\end{pmatrix}.
\label{eq:W3}
\end{equation}
Its eigenvalues $0,-4,-5$ are real and distinct, so by Lemma~\ref{lem:selfadjoint} $W$ is self-adjoint in some inner product. Its stationary state is $\pi=(1,1,2)/4$, every edge carries current $1/4$ around the cycle $1\to2\to3\to1$, the cycle affinity is $\ln(2\cdot3\cdot1)-\ln(1\cdot1\cdot1)=\ln6$, and $\sigma=(\kB/4)\ln6>0$.
\end{example}

Special spectra can nevertheless force detailed balance. Every two-state generator obeys it, and so does every diagonalizable generator with spectrum $\{0,-a,\dots,-a\}$, $a>0$: its spectral projector onto the eigenvalue $0$ is $\pi\mathbf{1}^{\T}$, so $W=-a(\mathbb{1}-\pi\mathbf{1}^{\T})$, and $W_{ij}\pi_j=a\pi_i\pi_j$ is symmetric. Spectral data therefore cannot certify detailed balance in general, although they can in special cases.

\subsection{Theorem}

\begin{theorem}[Spectral topology obstructs detailed balance but does not fix dissipation]
\label{thm:topodiss}
Under the assumptions above:
\begin{enumerate}
\item[(a)] \emph{Spectral obstruction.} If $W$ has a non-real eigenvalue, or is not diagonalizable (has an EP), then detailed balance fails, some cycle affinity is nonzero, and $\sigma>0$.
\item[(b)] \emph{Point-gap topology.} Let the graph be a ring of $N$ identical unit cells with translation-invariant rates of finite range $r$, $N>2r$, and Bloch generator $W(k)$. If $w(E)\neq0$ for some $E$ [Eq.~\eqref{eq:winding} with $\Lop=W$], then $\sigma>0$ for every such $N$.
\item[(c)] \emph{Quantitative bound.} For any time-integrated current $J_t=\sum_{i<j}d_{ij}N_{ij}(t)$ with antisymmetric weights, mean rate $j=\lim\langle J_t\rangle/t$ and diffusion coefficient $D=\lim\mathrm{Var}(J_t)/2t>0$,
\begin{equation}
\sigma\ \ge\ \kB\,\frac{j^2}{D}.
\label{eq:TUR}
\end{equation}
\item[(d)] \emph{Not a function of the winding number.} Within the class of (b), $\sigma$ is not a function of $w$: a family of generators with a fixed nonzero winding number can have $\sigma$ taking every value in $(0,\infty)$; in particular, $\sigma$ is not proportional to $w$.
\end{enumerate}
\end{theorem}

\begin{proof}
(a) By Proposition~\ref{prop:DB}(v) and Lemma~\ref{lem:selfadjoint}, detailed balance implies a real, diagonalizable spectrum. (b) If detailed balance held, translation invariance and uniqueness would make $\pi$ cell periodic, $\pi=(\pi_c,\dots,\pi_c)$, and the real-space matrix $\Pi^{-1/2}W\Pi^{1/2}$ would be symmetric and translation invariant. For $N>2r$ its Bloch symbol, $S(k)=\Pi_c^{-1/2}W(k)\Pi_c^{1/2}$, is then Hermitian for all real $k$, so $\det[W(k)-E]=\prod_n[\lambda_n(k)-E]$ with real $\lambda_n(k)$. For $E\notin\mathbb{R}$ no factor can wind around zero; for $E\in\mathbb{R}$ the determinant is real and nonvanishing and cannot wind either. Hence $w\equiv0$, a contradiction. (c) This is the thermodynamic uncertainty relation for steady-state currents~\cite{Barato2015,Gingrich2016,Pietzonka2016}; for the process started in its stationary state it holds at all finite times as $\mathrm{Var}(J_t)/\langle J_t\rangle^2\ge2\kB/(\sigma t)$~\cite{Horowitz2017,Horowitz2020}. (d) The Hatano-Nelson ring below has $w=-1$ throughout $\{p>q\}$ while $\sigma$ takes every value in $(0,\infty)$.
\end{proof}

Parts (a) and (b) are obstructions: they identify spectral features of a Markov generator that are incompatible with detailed balance. Example~\ref{ex:3state} shows that the converse fails: broken detailed balance need not produce complex spectra, EPs, or winding. Part (c) supplies a scale, but the scale comes from the current and its fluctuations, not from any invariant. Theorem~\ref{thm:topodiss} also does not establish that a bulk invariant protects a boundary current. Such correspondences have been demonstrated for particular classes of stochastic networks~\cite{Murugan2017,Dasbiswas2018,Tang2021}; Theorem~\ref{thm:topodiss} bounds the cost of any current that exists, Eq.~\eqref{eq:TUR}, but says nothing about its existence or direction.

Spectral data are not confined to obstruction once kinetic scales are included. Barato and Seifert conjectured that the number of coherent oscillations, set by the ratio of the imaginary to the real part of a leading eigenvalue, is bounded by the cycle affinity and the size of the network~\cite{Barato2017}. Kolchinsky, Ohga and Ito proved that the entropy production rate bounds how far the eigenvalues of a nonequilibrium generator can move from those of an equilibrium generator with the same stationary state and dynamical activity, so that fast oscillations and accelerated relaxation carry a thermodynamic cost~\cite{Kolchinsky2024}. These are quantitative statements about eigenvalue positions together with rate scales. Theorem~\ref{thm:topodiss}(d) concerns the integer winding number alone, which carries no rate scale and cannot fix $\sigma$.

Finally, $\pi$ is a rational function of the rates (Kirchhoff's construction of the stationary state~\cite{Schnakenberg1976}), so $\sigma$ is real-analytic in the rates on the positive orthant. An EP of $W$, where the spectral decomposition is singular, therefore leaves no singular signature in the stationary entropy production. Section~\ref{sec:EP} returns to this point.

\subsection{Exact example: the Hatano-Nelson ring}
\label{sec:HN}

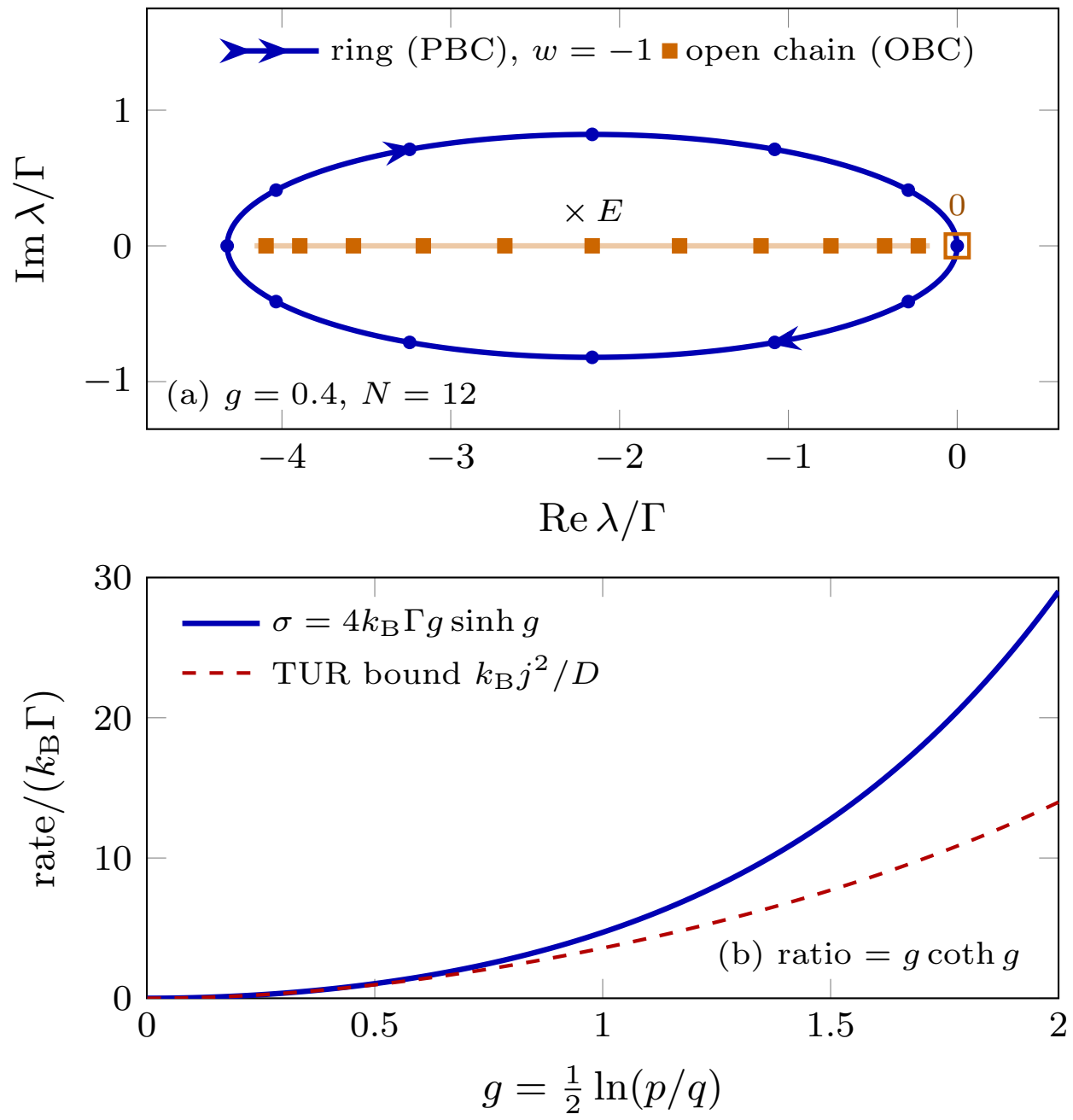
\begin{figure}[t]
\centering
\begin{tikzpicture}
\begin{groupplot}[group style={group size=1 by 2, vertical sep=1.15cm},
  width=\columnwidth, height=0.56\columnwidth,
  tick label style={font=\footnotesize}, label style={font=\small}]
\nextgroupplot[
  xlabel={$\mathrm{Re}\,\lambda/\Gamma$}, ylabel={$\mathrm{Im}\,\lambda/\Gamma$},
  xmin=-4.8, xmax=0.6, ymin=-1.35, ymax=1.75,
  legend style={font=\scriptsize, at={(0.5,0.98)}, anchor=north, draw=none, fill=none, legend columns=2},
  legend cell align=left]
\addplot[very thick, blue!70!black, domain=0:360, samples=181,
  postaction={decorate}, decoration={markings,
  mark=at position 0.14 with {\arrow{Stealth[length=2.5mm]}},
  mark=at position 0.64 with {\arrow{Stealth[length=2.5mm]}}}]
  ({2*cosh(0.4)*(cos(x)-1)}, {-2*sinh(0.4)*sin(x)});
\addlegendentry{ring (PBC), $w=-1$}
\addplot[only marks, mark=*, mark size=1.3pt, blue!70!black, samples at={0,30,...,330}, forget plot]
  ({2*cosh(0.4)*(cos(x)-1)}, {-2*sinh(0.4)*sin(x)});
\addplot[line width=1.2pt, orange!80!black, opacity=0.35, domain=0:1, samples=2, forget plot]
  ({-2*(cosh(0.4)+1) + x*4}, {0});
\addplot[only marks, mark=square*, mark size=1.5pt, orange!80!black, samples at={1,...,11}]
  ({-2*cosh(0.4)+2*cos(x*180/12)}, {0});
\addlegendentry{open chain (OBC)}
\addplot[only marks, mark=square, mark size=2.6pt, thick, orange!80!black, forget plot] coordinates {(0,0)};
\node[font=\scriptsize, anchor=south, orange!60!black] at (axis cs:0,0.12) {$0$};
\node[font=\scriptsize] at (axis cs:-2.16,0.25) {$\times\,E$};
\node[font=\scriptsize, anchor=south west] at (axis cs:-4.78,-1.33) {(a) $g=0.4$, $N=12$};
\nextgroupplot[
  xlabel={$g=\tfrac12\ln(p/q)$}, ylabel={rate$/(\kB\Gamma)$},
  xmin=0, xmax=2, ymin=0, ymax=30, domain=0.001:2, samples=101,
  legend style={font=\scriptsize, at={(0.02,0.97)}, anchor=north west, draw=none, fill=none},
  legend cell align=left]
\addplot[very thick, blue!70!black] {4*x*sinh(x)};
\addlegendentry{$\sigma=4\kB\Gamma g\sinh g$}
\addplot[thick, dashed, red!70!black] {4*sinh(x)*tanh(x)};
\addlegendentry{TUR bound $\kB j^2/D$}
\node[font=\scriptsize, anchor=east] at (axis cs:1.95,3) {(b) ratio $=g\coth g$};
\end{groupplot}
\end{tikzpicture}
\caption{\label{fig:HN}Hatano-Nelson ring. (a) Periodic spectrum, Eq.~\eqref{eq:HNspectrum}, traversed clockwise as $k$ increases (winding $w=-1$ about any interior point $E$; dots: $N=12$), and the real spectrum of the open chain with the same local rates and reflecting ends (squares, $N=12$): the stationary eigenvalue $0$ (open square) and $N-1$ eigenvalues $-2\Gamma\cosh g+2\Gamma\cos(m\pi/N)$ in the band shown shaded. The ring generator is normal, so the complex spectrum carries no EP. (b) Entropy production and the thermodynamic uncertainty bound. At fixed winding number the dissipation is continuous and unbounded; the ratio $g\coth g$ approaches one only in linear response.}
\end{figure}

Let a walker hop on a ring of $N$ sites with rates $p=\Gamma e^{g}$ clockwise and $q=\Gamma e^{-g}$ counterclockwise. The rate connection is uniform, $\Acon_{n+1,n}=2g$, so $g$, the ``imaginary vector potential'' of Hatano and Nelson~\cite{Hatano1996}, is exactly half the affinity per bond, and the ring holonomy is $\Acon[\text{ring}]=2Ng$. The Bloch generator is
\begin{equation}
\lambda(k)=pe^{-\ii k}+qe^{\ii k}-(p+q)=2\Gamma\big[\cos(k+\ii g)-\cosh g\big],
\label{eq:HNspectrum}
\end{equation}
an ellipse centred at $-2\Gamma\cosh g$ with semi-axes $2\Gamma\cosh g$ and $2\Gamma|\sinh g|$ [Fig.~\ref{fig:HN}(a)]. For $E$ inside the ellipse, $w(E)=-\sgn g$. The imaginary shift of $k$ is the spectral signature of the nonzero holonomy.

Because $W$ is circulant it is normal, and because $\pi$ is uniform the stationary inner product coincides with the standard one up to normalization. The ring is therefore normal, but not self-adjoint in the physical inner product; it has no EP for any $g$, yet its spectrum is complex for $g\neq0$. For $N=7$, $p=2$, $q=1$, for example, $[W,W^{\T}]=0$ exactly while $\sigma=\kB\ln2$.

The scaled cumulant generating function of the net number of clockwise jumps is $\theta(s)=p(e^s-1)+q(e^{-s}-1)$, giving $j=\theta'(0)=p-q=2\Gamma\sinh g$ and $D=\theta''(0)/2=(p+q)/2=\Gamma\cosh g$. With uniform $\pi$, Eq.~\eqref{eq:sigmapairing} gives $\sigma=\kB(p-q)\ln(p/q)=4\kB\Gamma g\sinh g$. Hence
\begin{equation}
\frac{\sigma}{\kB j^2/D}=g\coth g\ \ge 1,
\label{eq:gcothg}
\end{equation}
with equality only as $g\to0$, i.e., in linear response [Fig.~\ref{fig:HN}(b)]. Equation~\eqref{eq:gcothg} is the case of equality in the affinity-dependent bound of Pietzonka, Barato and Seifert~\cite{Pietzonka2016b}, $\sigma\ge\kB(j^2/D)(a/2)\coth(a/2)$, where $a$ is the smallest affinity per cycle length in the network; here $a/2=g$, and the uniform ring saturates the bound. Along the family $g>0$ the winding number is constant while $\sigma$ sweeps $(0,\infty)$; this proves Theorem~\ref{thm:topodiss}(d).

\begin{corollary}[Dissipation or accumulation]
\label{cor:skin}
On the open chain with the same local rates, the graph is a tree, every cycle affinity vanishes trivially, and the stationary state obeys detailed balance with $\pi_n\propto(p/q)^n=e^{2gn}$; then $\sigma=0$. The bulk winding number is unchanged and now appears as boundary accumulation, the stochastic skin effect, whose consequences for relaxation in one-dimensional stochastic processes are analyzed in Ref.~\cite{Sawada2024}.
\end{corollary}

The imaginary gauge transformation $W\mapsto e^{-gX}We^{gX}$ ($X$ the position operator) removes $g$ on the open chain, which is why the open-chain spectrum is real (with reflecting ends it is $\{0\}\cup\{-2\Gamma\cosh g+2\Gamma\cos(m\pi/N):m=1,\dots,N-1\}$), but cannot remove the holonomy $2Ng$ of the ring. Point-gap topology therefore forces either dissipation, when the relevant cycles close, or accumulation at a boundary, when they do not. In the two-dimensional network models of Refs.~\cite{Murugan2017,Dasbiswas2018,Tang2021} both occur together: the internal cycles carry nonzero affinity in every cell, so $\sigma>0$ throughout, while the boundary hosts a chiral current whose cost Eq.~\eqref{eq:TUR} bounds.

\emph{Torsion as a source of affinity.} Lattice defects can supply the cycle affinity that the open chain lacks. If the hopping bias of a walker is locked to the crystal frame rather than to the laboratory, a hop along lattice direction $\bm e_b$ has rate $\Gamma e^{\bm g\cdot\bm e_b}$ and its reverse $\Gamma e^{-\bm g\cdot\bm e_b}$, so $\Acon_b=2\bm g\cdot\bm e_b$. The holonomy around a closed loop of bonds is then $2\bm g\cdot\sum_b\bm e_b=2\bm g\cdot\bm b$, where $\bm b$ is the enclosed Burgers vector [the discrete form of Eq.~\eqref{eq:burgers}]; in continuum language, $\dd\Acon=2g_aT^a$. On a perfect lattice $\Acon$ is exact, $\pi\propto e^{2\bm g\cdot\bm y}$ obeys detailed balance, and an open sample shows only boundary accumulation (Corollary~\ref{cor:skin}). On an open sample without periodic directions every cycle of the bond graph is a sum of faces, so with dislocations present Proposition~\ref{prop:DB} implies that detailed balance holds if and only if $\bm g\cdot\bm b=0$ at every core (with periodic directions, the non-contractible cycles carry affinity $2\bm g$ paired with the period vectors, as on the ring, and must be counted too); each core with $\bm g\cdot\bm b\neq0$ carries the cycle affinity $2\bm g\cdot\bm b$ and is a source of entropy production. The defect-geometry and thermodynamic rows of Table~\ref{tab:dictionary} thus meet in a single model, whose quantitative consequences, including how the dissipation scales with sample size and bias, are worked out in a companion paper~\cite{Hunt2026dislocations}.

\subsection{Quantum generators}

Lemma~\ref{lem:selfadjoint} applies to Lindbladians as superoperators. Definitions of quantum detailed balance differ in how they treat a Hamiltonian part~\cite{Fagnola2010,Carlen2017,Tong2025}. If the whole generator is self-adjoint in the GNS or KMS inner product of the stationary state (strict symmetry), its spectrum is real and diagonalizable and it has no EPs~\cite{Carlen2017}. The notion due to Alicki and to Kossakowski, Frigerio, Gorini and Verri~\cite{Alicki1976,Kossakowski1977}, detailed balance up to a Hamiltonian part, permits in addition a Hamiltonian contribution that leaves the stationary state invariant and is anti-self-adjoint, the quantum counterpart of time-reversal-odd dynamics. It is this part that allows the Liouvillian EP of Sec.~\ref{sec:dephased}.

\section{Exceptional points and detailed balance}
\label{sec:EP}

\subsection{Normal form}

Near a twofold degeneracy any generator reduces to a $2\times2$ block, $\Lop=\bar\lambda\,\mathbb{1}+\left(\begin{smallmatrix}a&b\\ c&-a\end{smallmatrix}\right)$, with eigenvalues $\lambda_\pm=\bar\lambda\pm\sqrt{a^2+bc}$. An EP occurs where $a^2+bc=0$ while the traceless part is nonzero; that part is then nilpotent and $\Lop$ is a Jordan block. Near the EP, $\lambda_\pm-\bar\lambda\propto\pm\sqrt{\varepsilon}$ for a control parameter detuned by $\varepsilon$ (a Puiseux expansion~\cite{Kato1966}), the right and left eigenvectors become orthogonal, and the Petermann factor
\begin{equation}
K_n=\frac{\langle L_n|L_n\rangle\langle R_n|R_n\rangle}{|\langle L_n|R_n\rangle|^2}
\label{eq:petermann}
\end{equation}
diverges~\cite{Petermann1979,Wang2020}; here $\langle L_n|$ is the left eigenvector, a row vector with $\langle L_n|\Lop=\lambda_n\langle L_n|$, $|L_n\rangle$ its conjugate transpose, and $|R_n\rangle$ the right eigenvector. What is singular at an EP is the eigen-decomposition itself (modal amplitudes, Petermann factors); $e^{\Lop t}$ at fixed $t$ is analytic in the entries of $\Lop$, and in ordinary correlation and response functions of stable systems the divergent modal terms cancel. Encircling an EP in parameter space exchanges the two eigenvalues, the monodromy of the square-root branch~\cite{Heiss2012}; dynamically encircling it produces chiral state transfer~\cite{Xu2016}.

The matrix alone does not decide whether an EP is compatible with equilibrium. That requires the noise, or bath, and the behavior of each variable under time reversal.

\subsection{Generalized detailed balance with time-reversal parity}
\label{sec:GDB}

For linear Langevin dynamics, generator (e) of Sec.~\ref{sec:architecture}, assign each variable its parity under time reversal through $\Theta=\mathrm{diag}(\theta_i)$, $\theta_i=\pm1$ (positions even, momenta odd). The stationary process is invariant under $t\to-t$, $x\to\Theta x$, the physical definition of equilibrium, if and only if~\cite{Risken1989}
\begin{equation}
\Theta\Sigma\Theta=\Sigma,\qquad \Theta A\Theta\,\Sigma=\Sigma A^{\T}.
\label{eq:GDB}
\end{equation}
Equivalently, stationary correlation functions obey $\langle x_i(t)x_j(0)\rangle=\theta_i\theta_j\langle x_j(t)x_i(0)\rangle$, a condition that can be tested directly on measured trajectories. We call Eq.~\eqref{eq:GDB} generalized detailed balance (GDB). Define the stationary inner product $\langle x,y\rangle_\Sigma=x^{\T}\Sigma^{-1}y$; for a quadratic Hamiltonian $E(x)$ in thermal equilibrium, $\langle x,x\rangle_\Sigma=2E/\kB T$, so this is the energy norm of Sec.~\ref{sec:architecture} in thermal units.

\begin{proposition}[EPs and generalized detailed balance]
\label{prop:GDB}
Let $A$ be Hurwitz and $\Sigma>0$.
\begin{enumerate}
\item[(i)] If GDB holds, $A=A_e+A_o$ with $A_e=(A+\Theta A\Theta)/2=-\Xi\Sigma^{-1}$ self-adjoint and $A_o=(A-\Theta A\Theta)/2$ anti-self-adjoint in $\langle\cdot,\cdot\rangle_\Sigma$.
\item[(ii)] If GDB holds and all variables are even ($\Theta=\mathbb{1}$), then $A$ is self-adjoint in $\langle\cdot,\cdot\rangle_\Sigma$; its spectrum is real and it has no EP.
\item[(iii)] Hence, if all variables are even and $A$ has a non-real eigenvalue or an EP, GDB fails for every noise matrix $\Xi$; for $\Xi>0$ the entropy production $\sigma=\kB\tr(\Omega^{\T}\Xi^{-1}\Omega\Sigma)$, with $\Omega=A+\Xi\Sigma^{-1}$, is strictly positive.
\end{enumerate}
\end{proposition}

\begin{proof}
An operator $X$ is self-adjoint in $\langle\cdot,\cdot\rangle_\Sigma$ if and only if $X\Sigma$ is symmetric, and anti-self-adjoint if and only if $X\Sigma$ is antisymmetric. (i) Using $\Theta A\Theta\Sigma=\Sigma A^{\T}$, $A_e\Sigma=(A\Sigma+\Sigma A^{\T})/2=-\Xi$ is symmetric and $A_o\Sigma=(A\Sigma-\Sigma A^{\T})/2$ is antisymmetric. (ii) For $\Theta=\mathbb{1}$, $A_o=0$; apply Lemma~\ref{lem:selfadjoint}. (iii) The first claim is the contrapositive of (ii). The stationary probability current is $\Omega x\,p_{\mathrm{ss}}(x)$, and $\sigma=\kB\int J^{\T}\Xi^{-1}J/p_{\mathrm{ss}}$~\cite{Seifert2012} gives the stated trace; $\Omega=0$ if and only if $A\Sigma=-\Xi$, which is GDB with $\Theta=\mathbb{1}$, and otherwise $\sigma>0$ because $\Xi^{-1}>0$ and $\Sigma>0$.
\end{proof}

Proposition~\ref{prop:GDB} gives a checkable criterion. An EP compatible with equilibrium requires a nonzero time-reversal-odd part $A_o$, and it arises from the competition of $A_o$ with the dissipative part $A_e$, which the fluctuation-dissipation relation $A_e\Sigma=-\Xi$ ties to the noise. An EP of a system with only even variables certifies a nonequilibrium steady state. In neither case does the EP mark a threshold in stationary dissipation: $\Sigma$ solves a linear equation whose operator, $X\mapsto AX+XA^{\T}$, has eigenvalues $\lambda_i+\lambda_j\neq0$ for Hurwitz $A$, so $\Sigma$ and $\sigma$ are rational functions of the entries of $A$ throughout the stable region and are analytic across EPs. The same holds for Markov jump processes (Sec.~\ref{sec:thermo}). Only the spectral decomposition, and quantities built from it such as the Petermann factor, is singular at an EP.

\subsection{An EP in equilibrium: critical damping}
\label{sec:langevin}

Let $x=(q,p)$ with $\dd q=(p/m)\,\dd t$ and $\dd p=(-\kappa_s q-\gamma p)\,\dd t+\sqrt{2m\gamma\kB T}\,\dd W$, where $\kappa_s$ is the spring constant and $\gamma$ the damping rate. Then
\begin{equation}
\begin{gathered}
A=\begin{pmatrix}0&1/m\\-\kappa_s&-\gamma\end{pmatrix},\qquad \Xi=\begin{pmatrix}0&0\\0&m\gamma\kB T\end{pmatrix},\\
\Sigma=\kB T\begin{pmatrix}1/\kappa_s&0\\0&m\end{pmatrix},\qquad \Theta=\begin{pmatrix}1&0\\0&-1\end{pmatrix},
\end{gathered}
\label{eq:langevin}
\end{equation}
and GDB holds, with $A_e=\mathrm{diag}(0,-\gamma)$ the damping and $A_o=\left(\begin{smallmatrix}0&1/m\\-\kappa_s&0\end{smallmatrix}\right)$ the Hamiltonian flow. Treating all variables as even, GDB would fail, since $A\Sigma$ is not symmetric. The eigenvalues
\begin{equation}
\lambda_\pm=-\frac{\gamma}{2}\pm\sqrt{\frac{\gamma^2}{4}-\omega_0^2},\qquad \omega_0^2=\frac{\kappa_s}{m},
\label{eq:langevin-eigs}
\end{equation}
coalesce at critical damping, $\gamma=2\omega_0$, where $A$ is a Jordan block. Because the backward generator maps linear observables to linear observables through $A^{\T}$, the full Fokker-Planck generator is not diagonalizable at this point either. The asymptotic relaxation rate, $-\max\mathrm{Re}\,\lambda$, equals $\gamma/2$ below the EP and $\gamma/2-\sqrt{\gamma^2/4-\omega_0^2}$ above it, and is maximal, $\omega_0$, at the EP [Fig.~\ref{fig:EP}(a,b)]. This is the cleanest example of an EP in thermal equilibrium: nothing is driven, and the EP arises from the competition of the time-reversal-odd Hamiltonian flow with the time-reversal-even damping.

\subsection{The optomechanical dimer: same EP, driven realization}

Linearizing an optomechanical cavity driven on the red sideband and making the rotating-wave approximation gives a beam-splitter coupling $G$ between the intracavity field $a$ (energy decay rate $\kappa$) and the mechanical mode $b$ (rate $\gamma_m$)~\cite{Aspelmeyer2014}. The mean amplitudes obey
\begin{equation}
\frac{\dd}{\dd t}\begin{pmatrix}a\\ b\end{pmatrix}=\begin{pmatrix}-\kappa/2&\ii G\\ \ii G&-\gamma_m/2\end{pmatrix}\begin{pmatrix}a\\ b\end{pmatrix},
\end{equation}
with eigenvalues
\begin{equation}
\lambda_\pm=-\frac{\kappa+\gamma_m}{4}\pm\sqrt{\Big(\frac{\kappa-\gamma_m}{4}\Big)^2-G^2}.
\label{eq:OMeigs}
\end{equation}
For $\kappa\neq\gamma_m$ the EP is at $\GEP=|\kappa-\gamma_m|/4>0$, where the matrix is a Jordan block; for $\kappa=\gamma_m$ and $G=0$ the matrix is scalar, an ordinary degeneracy. For $G\ll\GEP$ and $\kappa>\gamma_m$ the mechanical energy decay rate becomes $\gamma_m+4G^2/(\kappa-\gamma_m)$, which reduces to the familiar optomechanical damping $4G^2/\kappa$ for $\kappa\gg\gamma_m$. For $G>\GEP$ the modes hybridize with normal-mode splitting $2\sqrt{G^2-\GEP^2}$; for $\gamma_m\to0$ the threshold $\GEP=\kappa/4$ is the known strong-coupling criterion~\cite{Dobrindt2008,Groblacher2009}. Below the EP the Petermann factor of either mode is
\begin{equation}
K=\frac{\GEP^2}{\GEP^2-G^2},\qquad G<\GEP ,
\label{eq:OMpetermann}
\end{equation}
diverging at the EP. EP encircling has been demonstrated on a related optomechanical platform, between two mechanical modes coupled through a common optical cavity~\cite{Xu2016}.

Is this EP an equilibrium one? The matrix does not say. Within the rotating-wave approximation, the same amplitude equations, with $G=k_c/(2m\omega_0)$, describe two degenerate resonators coupled by a spring $k_c$, each damped by its own bath; with both baths at one temperature and $|k_c|<m\omega_0^2$, that lab-frame system obeys GDB with momenta odd (verified in Appendix~\ref{app:EP}). The EP therefore requires neither drive nor nonreciprocity. The optomechanical realization is nonetheless driven: the beam-splitter coupling is induced by a red-detuned laser, and the steady state carries heat from the mechanical bath to the effectively cold optical bath, which is the mechanism of sideband cooling~\cite{Aspelmeyer2014}. The EP and the Petermann factor are properties of the mean-field dynamics in both cases; whether the steady state is an equilibrium one is a separate question about the baths and the drive.

\subsection{A Liouvillian EP at infinite temperature: the dephased dimer}
\label{sec:dephased}

Consider two sites coupled by tunnelling, $H=\Delta(|1\rangle\langle2|+|2\rangle\langle1|)$, with pure dephasing $L_k=\sqrt{\gamma_\phi}\,|k\rangle\langle k|$, $k=1,2$, so that coherences decay at rate $\gamma_\phi$. The Liouvillian spectrum is
\begin{equation}
\Big\{0,\ -\gamma_\phi,\ \tfrac12\big(-\gamma_\phi\pm\sqrt{\gamma_\phi^2-16\Delta^2}\big)\Big\}.
\label{eq:HSspectrum}
\end{equation}
The Bloch components obey $\dot x=-\gamma_\phi x$, $\dot y=-\gamma_\phi y-2\Delta z$, $\dot z=2\Delta y$, so the population sector has characteristic equation $\lambda^2+\gamma_\phi\lambda+4\Delta^2=0$: exactly that of the Langevin oscillator, Eq.~\eqref{eq:langevin-eigs}, with $\omega_0=2\Delta$ and $\gamma=\gamma_\phi$. A Liouvillian EP, with a Jordan block of the full Liouvillian, sits at $\gamma_\phi=4\Delta$. Below it, populations oscillate; above it they relax monotonically, and for $\gamma_\phi\gg\Delta$ the slow rate tends to $4\Delta^2/\gamma_\phi$, incoherent hopping at $k_{\mathrm{hop}}=2\Delta^2/\gamma_\phi$, the Haken-Strobl limit~\cite{Haken1973}.

The Liouvillian gap, the asymptotic relaxation rate, is $\gamma_\phi/2$ for $\gamma_\phi\le4\Delta$ and $(\gamma_\phi-\sqrt{\gamma_\phi^2-16\Delta^2})/2$ above, and is therefore maximal, $2\Delta$, at the EP, the exact analogue of critical damping. This is a statement about asymptotic relaxation only. The time of first population transfer is shortest in the coherent limit $\gamma_\phi\to0$, and the transfer efficiency to a sink, the figure of merit in experiments on dephasing-assisted transport~\cite{Maier2019}, is a different quantity again.

This example also separates two notions of quantum EP~\cite{Minganti2019}. The no-jump effective Hamiltonian $H_{\mathrm{eff}}=H-\tfrac{\ii}{2}\sum_kL_k^\dagger L_k=H-\tfrac{\ii}{2}\gamma_\phi\mathbb{1}$ has eigenvalues $\pm\Delta-\ii\gamma_\phi/2$ and \emph{no} EP for any $\gamma_\phi$. The EP of Eq.~\eqref{eq:HSspectrum} is created entirely by the quantum-jump terms, consistent with the decoherence-induced Liouvillian EPs observed in a superconducting qubit~\cite{Chen2022}.

Zero stationary current would not by itself show that this Liouvillian describes equilibrium. Its structure does. In the Hilbert-Schmidt inner product the dissipator, built from Hermitian jump operators, is self-adjoint, and the Hamiltonian part $-\ii[H,\cdot\,]$ is anti-self-adjoint; since the stationary state is maximally mixed, the Hilbert-Schmidt product is the stationary (GNS) one. The dimer therefore satisfies quantum detailed balance in the sense of Alicki and of Kossakowski et al.~\cite{Alicki1976,Kossakowski1977} with respect to the infinite-temperature state, and its EP has the same origin as at critical damping. For this site-dephased hopping model the stationary state is maximally mixed, so the statement concerns infinite temperature; a statement at finite temperature would require a bath model with a nontrivial stationary state.

\subsection{An EP that certifies nonequilibrium: odd elasticity}

Take the two-dimensional isotropic constitutive law with bulk modulus $B$, shear modulus $\mu$ and odd shear modulus $K^o$~\cite{Scheibner2020,Fruchart2023},
\begin{equation}
\sigma_{ij}=B\,\delta_{ij}\nabla\cdot\bu+\mu\big(s_1\tau^z_{ij}+s_2\tau^x_{ij}\big)+K^o\big(s_2\tau^z_{ij}-s_1\tau^x_{ij}\big),
\end{equation}
with shear strains $s_1=\partial_xu_x-\partial_yu_y$, $s_2=\partial_xu_y+\partial_yu_x$ and Pauli matrices $\tau^{x,z}$. The odd term maps each shear strain into the other shear stress antisymmetrically, violating the major symmetry $C_{ijkl}=C_{klij}$. With overdamped dynamics $\zeta\partial_t\bu=\nabla\cdot\sigma$, the dynamical matrix in the basis of longitudinal and transverse components is
\begin{equation}
D(\bk)=k^2\begin{pmatrix}B+\mu&K^o\\ -K^o&\mu\end{pmatrix},
\label{eq:oddD}
\end{equation}
independent of the direction of $\bk$, with relaxation rates $\zeta^{-1}k^2\big[B/2+\mu\pm\sqrt{B^2/4-(K^o)^2}\big]$. For $B>0$ and $k\neq0$ the EP is at $|K^o|=B/2$; beyond it the rates are complex and the overdamped medium supports odd elastic waves, in agreement with Scheibner et al.~\cite{Scheibner2020} (whose condition $K^oA+(K^o)^2>(B/2)^2$ reduces to ours when the odd modulus $A$ vanishes).

\begin{figure*}[t]
\centering
\begin{tikzpicture}
\begin{groupplot}[group style={group size=3 by 1, horizontal sep=1.55cm},
  width=0.335\textwidth, height=0.27\textwidth,
  tick label style={font=\footnotesize}, label style={font=\small}]
\nextgroupplot[xlabel={$\gamma/\omega_0$}, ylabel={$-\mathrm{Re}\,\lambda_\pm/\omega_0$},
  xmin=0, xmax=4, ymin=0, ymax=4, domain=0:4, samples=401]
\addplot[thick, blue!70!black] {x/2 + sqrt(max(0,x^2/4-1))};
\addplot[thick, red!70!black] {x/2 - sqrt(max(0,x^2/4-1))};
\addplot[line width=2.2pt, black, opacity=0.35] {x/2 - sqrt(max(0,x^2/4-1))};
\addplot[thin, dashed, black] coordinates {(2,0) (2,4)};
\node[font=\scriptsize, anchor=west] at (axis cs:2.05,3.7) {EP};
\node[font=\scriptsize, anchor=north west] at (axis cs:0.05,3.95) {(a) decay rates};
\nextgroupplot[xlabel={$\gamma/\omega_0$}, ylabel={$\mathrm{Im}\,\lambda_\pm/\omega_0$},
  xmin=0, xmax=4, ymin=-1.15, ymax=1.15, domain=0:4, samples=401]
\addplot[thick, blue!70!black] {sqrt(max(0,1-x^2/4))};
\addplot[thick, red!70!black] {-sqrt(max(0,1-x^2/4))};
\addplot[thin, dashed, black] coordinates {(2,-1.15) (2,1.15)};
\node[font=\scriptsize, anchor=south east] at (axis cs:3.95,-1.12) {(b) frequencies};
\nextgroupplot[xlabel={$K^o/B$}, ylabel={rate$\,\times\,\zeta/(k^2B)$},
  xmin=0, xmax=1, ymin=0, ymax=3.1, domain=0:1, samples=401,
  legend style={font=\scriptsize, at={(0.02,0.98)}, anchor=north west, draw=none, fill=none},
  legend cell align=left]
\addplot[thick, blue!70!black] {1.5 + sqrt(max(0,0.25-x^2))};
\addlegendentry{$-\mathrm{Re}\,\lambda_\pm$}
\addplot[thick, red!70!black, forget plot] {1.5 - sqrt(max(0,0.25-x^2))};
\addplot[very thick, dashed, green!50!black] {4*x^2/3};
\addlegendentry{$\sigma_{\bk}/\kB$}
\addplot[thin, dashed, black, forget plot] coordinates {(0.5,0) (0.5,3.1)};
\node[font=\scriptsize, anchor=north east] at (axis cs:0.98,2.45) {(c) odd elastic mode};
\end{groupplot}
\end{tikzpicture}
\caption{\label{fig:EP}Exceptional points with and without detailed balance. (a,b) Langevin oscillator in thermal equilibrium, Eq.~\eqref{eq:langevin-eigs}: decay rates and frequencies versus damping. The asymptotic relaxation rate (grey band) is maximal at the EP, critical damping $\gamma=2\omega_0$. The population sector of the dephased dimer, Eq.~\eqref{eq:HSspectrum}, is described by the same curves with $\omega_0=2\Delta$ and $\gamma=\gamma_\phi$. (c) Odd elastic standing-wave mode with thermal noise, Eqs.~\eqref{eq:oddD} and \eqref{eq:oddsigma}, for $\mu=B$: the relaxation rates coalesce at $|K^o|=B/2$ and acquire imaginary parts beyond it, while the entropy production is analytic through the EP.}
\end{figure*}
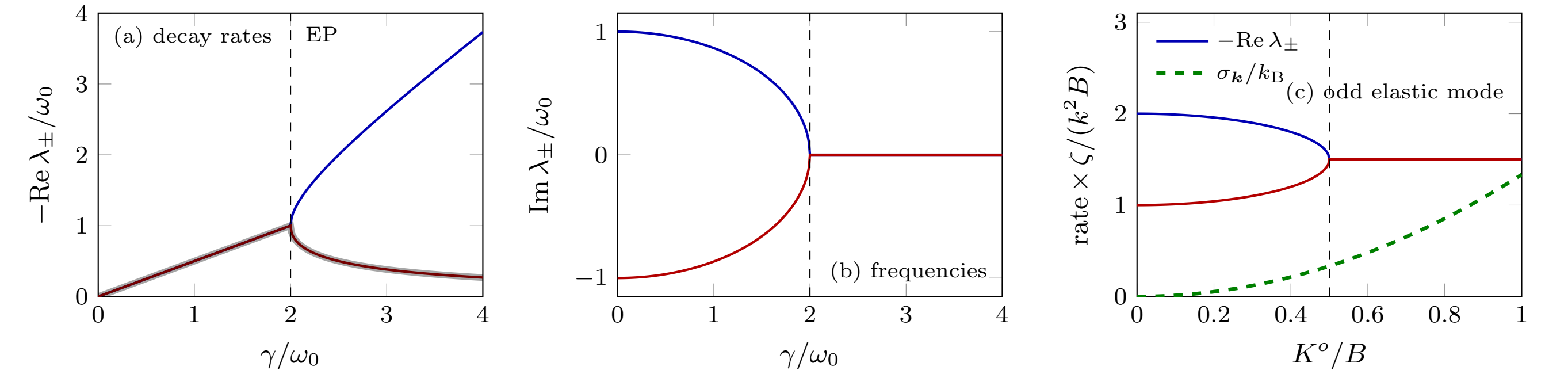

Displacements are even under time reversal. Without noise, $\zeta^{-1}D(\bk)$ is self-adjoint in $\langle x,y\rangle_\zeta=x^\dagger\zeta y$ if and only if $D(\bk)$ is Hermitian, i.e., $K^o=0$; an EP therefore certifies broken major symmetry, so the stress does not derive from an elastic energy and net work can be extracted over closed strain cycles~\cite{Scheibner2020}, which requires an energy source. With thermal noise obeying the Einstein relation, $\Xi=(\kB T/\zeta)\mathbb{1}$ for the amplitudes $(u_L,u_T)$ of a standing-wave mode, Proposition~\ref{prop:GDB}(iii) applies: GDB fails whenever $K^o\neq0$, and solving for $\Sigma$ (Appendix~\ref{app:EP}) gives
\begin{equation}
\sigma_{\bk}=\frac{4\kB k^2}{\zeta}\,\frac{(K^o)^2}{B+2\mu}.
\label{eq:oddsigma}
\end{equation}
The entropy production is quadratic in $K^o$, independent of temperature, and analytic through the EP [Fig.~\ref{fig:EP}(c)]. The EP changes the character of relaxation, from monotonic to oscillatory, with no signature in dissipation, just as the winding number of Sec.~\ref{sec:HN} leaves $\sigma$ free. Realizations include robotic micropolar metamaterials~\cite{Chen2021odd} and living chiral crystals~\cite{Tan2022}, whose fluctuations are not thermal, so Eq.~\eqref{eq:oddsigma} applies to them only as a model. More broadly, in non-reciprocal many-body systems, EPs at which a Goldstone mode coalesces with a damped mode mark the onset of time-dependent chiral phases~\cite{Fruchart2021}.

\subsection{A remark on exceptional-point metrology}

The square-root response near an EP was proposed as a route to enhanced sensing~\cite{Wiersig2014}, and enhanced splittings were observed~\cite{Chen2017,Hodaei2017}. The same eigenvector coalescence that produces the square root, however, also amplifies noise: in lasers and other active resonators the Petermann factor \eqref{eq:petermann} multiplies the fundamental linewidth~\cite{Petermann1979,Wang2020}, and analyses of signal-to-noise ratio find no generic advantage of passive EP sensors~\cite{Langbein2018,Lau2018,Wang2020}. For the optomechanical dimer, Eq.~\eqref{eq:OMpetermann} gives the eigenvector non-orthogonality explicitly; turning it into a measured excess-noise factor requires the noise covariance, the input-output relations and the observable, which we do not develop. Any claimed EP-enhanced sensitivity should therefore be reported as a signal-to-noise ratio, not as a responsivity.

\section{Benchtop tests}
\label{sec:experiments}

Each statement above is an identity or a theorem, so an experiment cannot refute it within its assumptions; it can, however, find the assumptions violated (for example, hidden slow variables that make the dynamics non-Markovian), and it can test the stated quantitative predictions. We list one platform per statement.

\emph{Odd-viscous Chern numbers.} Chiral fluids of spinning colloids~\cite{Soni2019} allow $\Omega$ and $\nu_o$ to be extracted from the dynamic structure factor $S(\bk,\omega)$: the gap at $k=0$ gives $|\Omega|$ and the curvature of $\omega_\pm(k)$ gives $\nu_o$. Equation~\eqref{eq:chern} predicts a change of $C_\pm$ from $0$ to $\pm2$ when the relative sign of $\Omega$ and $\nu_o$ is reversed, for instance by counter-rotating the frame. The accompanying change of edge spectra should be compared with calculations for the experimental boundary condition, not with bulk Chern numbers alone~\cite{Tauber2020,Graf2021}.

\emph{Torsion-topology pairing.} Laser-cut or three-dimensionally printed deformed kagome lattices~\cite{Paulose2015,Rocklin2017} with engineered dislocation dipoles allow the Burgers vector to be measured by a Burgers circuit on images and the localized floppy modes to be counted by indentation or vibrational spectroscopy. Equation~\eqref{eq:torsion-count} predicts the topological part of the count, its sign reversal at the two members of a dipole, and its exchange when $\PT$ is reversed by a Guest-Hutchinson deformation; disorder that preserves $\PT$ should leave the count unchanged.

\emph{Topology-dissipation bound.} A colloid driven around a toroidal optical trap with a periodic potential~\cite{Blickle2007} realizes the Hatano-Nelson ring in the deep-well (hopping) limit. Forward and backward jump counts give $p$, $q$ and hence $g$; trajectories give $j$ and $D$; the heat flow gives $\sigma$. Equation~\eqref{eq:gcothg} predicts the ratio $g\coth g$ with no free parameters. Opening the ring by a barrier converts the current into the boundary accumulation $\pi_n\propto e^{2gn}$ of Corollary~\ref{cor:skin}, at the same local rates. A three-well ring with rates tuned to Eq.~\eqref{eq:W3} would show directly that all measured relaxation rates can be real while a cycle current flows. Single-electron counting in quantum dots~\cite{Kung2012} offers a second, electronic platform with direct access to full counting statistics. Deterministic non-reciprocal metamaterials, robotic~\cite{Brandenbourger2019,Ghatak2020} and electrical~\cite{Helbig2020}, display the corresponding skin effect; there the injected power is the natural dissipation proxy, but Eq.~\eqref{eq:TUR}, a statement about stochastic currents, does not apply.

\emph{EPs compatible with equilibrium.} An optically trapped microsphere in gas, whose ballistic Brownian motion can be resolved~\cite{Li2010} and whose damping rate is set by the gas pressure, crosses critical damping in thermal equilibrium. The EP appears as the coalescence of the two poles of the position power spectral density and as the maximum of the relaxation rate of the mean, Eq.~\eqref{eq:langevin-eigs}; GDB can be tested on the same trajectories with the correlation-function criterion of Sec.~\ref{sec:GDB}, using $\theta_p=-1$. For the dephased dimer, trapped ions~\cite{Maier2019} or transmons with engineered dephasing~\cite{Chen2022} can resolve the change from oscillatory to monotonic population transfer at $\gamma_\phi=4\Delta$, the maximum of the Liouvillian gap there, and the $2\Delta^2/\gamma_\phi$ hopping rate above it. In optomechanics, sweeping the drive power moves $G$ through $\GEP$~\cite{Groblacher2009}; ringdown or optomechanically induced transparency spectra yield $\lambda_\pm$, and, with a model of the noise and of the input-output relations, noise spectra constrain the eigenvector non-orthogonality of Eq.~\eqref{eq:OMpetermann}. Comparing with a passively coupled pair of resonators at a common temperature would separate the EP from the drive.

\emph{EPs that certify nonequilibrium.} In odd elastic media~\cite{Chen2021odd,Tan2022}, relaxation spectra measured at fixed $k$ should show the coalescence of Eq.~\eqref{eq:oddD} at $|K^o|=B/2$, with the power input required to maintain $K^o$ measured independently. Where fluctuations are thermal, the correlation asymmetry $\langle u_L(t)u_T(0)\rangle\neq\langle u_T(t)u_L(0)\rangle$ measures the violation of GDB directly (time-correlation functions of odd Langevin systems are analysed in Ref.~\cite{Yasuda2022}), and Eq.~\eqref{eq:oddsigma} predicts an entropy production that grows as $(K^o)^2$ with no feature at the EP.

A common protocol serves all EP experiments: reconstruct the $2\times2$ generator block from response or ringdown data and evaluate its discriminant to locate the EP; then determine the time-reversal parities of the variables and test GDB with the measured noise, through the correlation-function criterion. The first step characterizes the deterministic dynamics; only the second says anything about equilibrium.

\section{Discussion}
\label{sec:discussion}

\begin{table*}[t]
\caption{\label{tab:systems}Summary of the worked examples. ``Detailed balance'' refers to the physically distinguished inner product and, where relevant, the time-reversal parities of the variables.}
\begin{ruledtabular}
\footnotesize
\begin{tabular}{p{3.0cm}p{2.7cm}p{3.3cm}p{3.3cm}p{4.3cm}}
\raggedright System &\raggedright Generator &\raggedright Spectrum and EPs &\raggedright Holonomy or invariant &\raggedright Detailed balance\tabularnewline
\colrule
\raggedright Odd-viscous fluid, Eq.~\eqref{eq:Hodd} &\raggedright $-\ii H(\bk)$, $H$ Hermitian &\raggedright frequencies (spectrum of $H$) real; no EP &\raggedright $C_m=m(\sgn\Omega+\sgn\nu_o)$ &\raggedright conservative (no bath)\tabularnewline
\raggedright Maxwell lattice with dislocation &\raggedright $\ii\Lop$ Hermitian &\raggedright frequencies (spectrum of $\ii\Lop$) real; no EP &\raggedright $\nu_T=\PT\cdot R_{\pi/2}\!\int\! \bm T/\Vc$ &\raggedright conservative (no bath)\tabularnewline
\raggedright Hatano-Nelson ring &\raggedright rate matrix; normal &\raggedright complex for $g\neq0$; no EP; $w=-\sgn g$ &\raggedright affinity $2Ng$ &\raggedright broken; $\sigma=4\kB\Gamma g\sinh g$\tabularnewline
\raggedright Three-state network, Eq.~\eqref{eq:W3} &\raggedright rate matrix &\raggedright real, distinct; no EP &\raggedright affinity $\ln6$ &\raggedright broken; $\sigma=(\kB/4)\ln6$\tabularnewline
\raggedright Langevin oscillator &\raggedright Langevin drift &\raggedright EP at $\gamma=2\omega_0$ &\raggedright (none) &\raggedright holds, with momentum odd\tabularnewline
\raggedright Optomechanical dimer &\raggedright mean-field amplitudes &\raggedright EP at $G=\GEP$ &\raggedright (none) &\raggedright realization driven; same matrix arises in equilibrium\tabularnewline
\raggedright Dephased dimer &\raggedright Lindbladian &\raggedright EP at $\gamma_\phi=4\Delta$ &\raggedright (none) &\raggedright holds at infinite temperature (Alicki; Kossakowski et al.)\tabularnewline
\raggedright Odd elastic mode &\raggedright Langevin drift; even variables &\raggedright EP at $|K^o|=B/2$ &\raggedright (none) &\raggedright broken for $K^o\neq0$; $\sigma_{\bk}$ analytic at EP\tabularnewline
\end{tabular}
\end{ruledtabular}
\end{table*}

Table~\ref{tab:systems} summarizes the examples. What is genuinely common is mathematical: connections whose holonomies are observable, index theorems that tie bulk holonomies to boundary or defect modes, and the spectral theory of generators that are not self-adjoint in the physical inner product. These structures recur because the systems share linearity, translation invariance or a graph structure, and the presence or absence of time-reversal symmetry; they do not recur because the systems share microscopic physics. The odd-viscous fluid carries a Chern number, and the Maxwell lattice a winding (polarization) invariant, for the same structural reasons as electrons in a magnetic field and in polarized insulators do, and for no deeper one.

Three points deserve emphasis. First, the precise content of the loose slogan ``non-Hermiticity is nonequilibrium'' is this: for stochastic dynamics with even variables, failure of self-adjointness in the stationary inner product is broken detailed balance. Self-adjointness in some other inner product, which a real spectrum provides, does not imply detailed balance (Example~\ref{ex:3state}), and normality is irrelevant (the Hatano-Nelson ring). The Hatano-Nelson imaginary vector potential is half the affinity per bond, and entropy production is the pairing of current with holonomy, Eq.~\eqref{eq:sigmapairing}. Second, spectral topology constrains dissipation only as an obstruction. A universal relation between entropy production and the winding number is excluded by Theorem~\ref{thm:topodiss}(d): an integer invariant carries no rate scale, so any relation must involve model-specific kinetic scales, as the quantitative spectral bounds of Refs.~\cite{Barato2017,Kolchinsky2024} do. Whether a protected boundary current exists is a model-specific question that Theorem~\ref{thm:topodiss} does not settle. Third, EPs require a separate assessment of the dynamics. With time-reversal-odd variables they are compatible with equilibrium (critical damping; the dephased dimer at infinite temperature); with only even variables they certify a nonequilibrium steady state (odd elasticity); and in neither case do they mark a threshold in stationary dissipation.

Several questions remain open. (i) Can the magnitude of a protected stochastic edge current be bounded from above or below in terms of the bulk gap of $W$ and the cycle affinities, sharpening Eq.~\eqref{eq:TUR} for topological networks? (ii) How do torsion (dislocations) and point-gap topology combine in active elastic media, where dislocations could bind skin modes~\cite{Schindler2021,Bhargava2021} while odd moduli supply non-reciprocity? For lattice-biased random walks the thermodynamic criterion $\bm g\cdot\bm b\neq0$ and the spectral criterion for the dislocation skin effect turn out to be complementary~\cite{Hunt2026dislocations}; whether a similar division holds in elastic media is open. (iii) Stationary quantities are analytic across EPs, and so are the propagator at fixed time and ordinary correlation and response functions; what is singular is the eigen-decomposition. Is there a sharp bound relating EP proximity to the precision of finite-time currents? (iv) For periodically driven stochastic pumps, whose geometric currents are linear in the driving frequency, the steady-state TUR is replaced by generalized relations~\cite{Koyuk2020}; what, if anything, do these imply for the cost of geometric pumping? Each is a well-posed problem within the language developed here.

\begin{acknowledgments}
\emph{Use of AI tools.} The author, who is responsible for its content, initiated and directed this work, set its scope, and organized independent critical review of its drafts. Claude (Anthropic; Claude Opus~5.5) was used for derivations, symbolic and numerical checks (Wolfram Language and Python; scripts provided as Supplemental Material), literature searches, and drafting of the text. ChatGPT (OpenAI; GPT-6 Astra) was used in the author's review of earlier drafts; that review supplied Example~\ref{ex:3state} and the normality counterexample. In a final review it reran the Python checks, of the preceding version and of this one, in a separate computing environment (all passed) and identified overstated claims, missing qualifications and missing references, which are corrected in this version.
The author reran the Python checks, in their present form, on a separate computer, and all passed.
\end{acknowledgments}

\appendix

\section{Spin-1 structure and Chern numbers of the odd-viscous fluid}
\label{app:spin1}

With $H(\bk)=ck_x\Lambda_1+ck_y\Lambda_2+f(k)\Lambda_3$ from Eq.~\eqref{eq:Hodd},
\begin{equation}
\Lambda_1=\begin{pmatrix}0&1&0\\1&0&0\\0&0&0\end{pmatrix},\
\Lambda_2=\begin{pmatrix}0&0&1\\0&0&0\\1&0&0\end{pmatrix},\
\Lambda_3=\begin{pmatrix}0&0&0\\0&0&\ii\\0&-\ii&0\end{pmatrix},
\end{equation}
one verifies $[\Lambda_a,\Lambda_b]=-\ii\epsilon_{abc}\Lambda_c$ and $\sum_a\Lambda_a^2=2\cdot\mathbb{1}$. Hence $S_a=-\Lambda_a$ is a spin-1 representation, $H=-\bm d\cdot\bm S$ with $\bm d=(ck_x,ck_y,f)$, and the eigenvalues are $0,\pm|\bm d|$. For a spin-$s$ Hamiltonian $\bm d\cdot\bm S$, the eigenstate with $\hat{\bm d}\cdot\bm S=m_s$ has Berry curvature equal to $-m_s$ times the solid-angle density of $\hat{\bm d}$; the overall sign depends on orientation conventions, which we fix by direct evaluation of Eq.~\eqref{eq:berry}. By rotational symmetry the curvature depends only on $k$, and $C_n=\int_0^\infty k\,\Fcurv_{n,xy}(k,0)\,\dd k$. Numerical integration (relative accuracy $10^{-6}$) with $c=1$ gives:
\begin{center}
\begin{tabular}{cccc}
\toprule
$(\Omega,\nu_o)$ & $C_-$ & $C_0$ & $C_+$\\
\midrule
$(+1,+0.5)$ & $-2$ & $0$ & $+2$\\
$(+1,-0.5)$ & $0$ & $0$ & $0$\\
$(-1,+0.5)$ & $0$ & $0$ & $0$\\
$(-1,-0.5)$ & $+2$ & $0$ & $-2$\\
\midrule
$(+1,0)$ & $-1$ & $0$ & $+1$\\
$(-1,0)$ & $+1$ & $0$ & $-1$\\
$(+1,+0.001)$ & $-2$ & $0$ & $+2$\\
\bottomrule
\end{tabular}
\end{center}
The first four rows agree with Eq.~\eqref{eq:chern}. The last three show that for $\nu_o=0$ the integrated curvature is $\pm\sgn\Omega$, a hemisphere's worth, and that an odd viscosity three orders of magnitude smaller than in the first rows restores the quantized value. The identity $\nabla^2\bv^*=\nabla\omega$ for incompressible flow follows by writing $\bv=(\partial_y\psi,-\partial_x\psi)$, so that $\omega=-\nabla^2\psi$ and $\bv^*=-\nabla\psi$.

\section{Markov-network details}
\label{app:HN}

\emph{Hatano-Nelson ring.} With $p_n$ the probability of site $n$, $\dot p_n=p\,p_{n-1}+q\,p_{n+1}-(p+q)p_n$, and $p_n\propto e^{\ii kn}$ gives Eq.~\eqref{eq:HNspectrum}, using $pe^{-\ii k}+qe^{\ii k}=2\Gamma\cos(k+\ii g)$. Writing $\lambda(k)-E_0=pe^{-\ii k}+qe^{\ii k}$ for the centre $E_0=-(p+q)$, the term $pe^{-\ii k}$ dominates when $p>q$, so the curve winds once clockwise and $w(E_0)=-1$; numerical evaluation of Eq.~\eqref{eq:winding} gives $w=-1,+1,-1$ for $(p,q)=(2,1),(1,2),(3,0.5)$. The counting-field-tilted generator for the net clockwise jump number is $\theta(s)=p(e^s-1)+q(e^{-s}-1)$. The inequality $g\coth g\ge1$ is equivalent, with $x=p/q$, to $\ln x\ge2(x-1)/(x+1)$ for $x>1$. On the open chain of $L$ sites with reflecting ends, the null vector of $W$ is $\pi_n\propto(p/q)^n$. For $N=7$, $p=2$, $q=1$, the commutator $[W,W^{\T}]$ vanishes identically, six of the seven eigenvalues are non-real, and Eq.~\eqref{eq:sigma} gives $\sigma=\kB\ln2$.

\emph{Three-state network.} For Eq.~\eqref{eq:W3}, the eigenvalues are $0,-4,-5$; with $V$ the matrix of right eigenvectors, $G=(VV^{\T})^{-1}$ is positive definite and satisfies $GW=W^{\T}G$. The null vector is $\pi=(1,1,2)/4$, $W\,\mathrm{diag}(\pi)$ is not symmetric, the forward and backward cycle-rate products are $6$ and $1$, and $\sigma=(\kB/4)\ln6$.

\section{Exceptional-point computations}
\label{app:EP}

\emph{Langevin oscillator.} With Eq.~\eqref{eq:langevin}, $A\Sigma+\Sigma A^{\T}+2\Xi=0$ and $\Theta A\Theta\Sigma=\Sigma A^{\T}$ hold identically, while $A\Sigma-\Sigma A^{\T}=2\kB T\left(\begin{smallmatrix}0&1\\-1&0\end{smallmatrix}\right)\neq0$. At $\gamma=2\omega_0$ the Jordan form of $A$ is $\left(\begin{smallmatrix}-\omega_0&1\\0&-\omega_0\end{smallmatrix}\right)$. Numerical maximization of $-\max\mathrm{Re}\,\lambda$ over $\gamma$ (with $m=\kappa_s=1$) returns $\gamma=2$, rate $1$.

\emph{Coupled resonators at a common temperature.} For $H=\sum_j[p_j^2/2m+m\omega_0^2q_j^2/2]+k_cq_1q_2$ with damping rates $\gamma_{1,2}$ on the momenta and noise $\Xi=\kB T\,\mathrm{diag}(0,\gamma_1m,0,\gamma_2m)$, the Gibbs covariance $\Sigma=\kB T\,(\nabla^2H)^{-1}$ solves the Lyapunov equation and satisfies Eq.~\eqref{eq:GDB} with $\Theta=\mathrm{diag}(1,-1,1,-1)$, for $|k_c|<m\omega_0^2$ (so that $H$ is bounded below and $\Sigma$ is positive definite) and $\gamma_1,\gamma_2>0$. For $|k_c|\ge m\omega_0^2$ the Hessian of $H$ is not positive definite and there is no Gibbs state.

\emph{Optomechanical dimer.} The discriminant of the characteristic polynomial vanishes at $G=(\kappa-\gamma_m)/4$ for $\kappa>\gamma_m$, where the Jordan form is
\begin{equation}
J_{\mathrm{EP}}=\begin{pmatrix}-(\kappa+\gamma_m)/4&1\\0&-(\kappa+\gamma_m)/4\end{pmatrix}.
\end{equation}
Expanding the slow root, $\lambda_+=-\gamma_m/2-2G^2/(\kappa-\gamma_m)+O(G^4)$. With right eigenvectors $|R_\pm\rangle$ and left eigenvectors $\langle L_\pm|$ (rows; eigenvectors of the transpose), and $|L_\pm\rangle$ the conjugate transposes of the latter, Eq.~\eqref{eq:petermann} simplifies to $K=(\kappa-\gamma_m)^2/[(\kappa-\gamma_m)^2-16G^2]$, which is Eq.~\eqref{eq:OMpetermann}.

\emph{Dephased dimer.} In the vectorized representation, $\hat\Lop=\hat\Lop_H+\hat\Lop_D$ with $\hat\Lop_H=-\ii(\mathbb{1}\otimes H-H^{\T}\otimes\mathbb{1})$ and $\hat\Lop_D=\sum_k\big[L_k^*\otimes L_k-\tfrac12\mathbb{1}\otimes L_k^\dagger L_k-\tfrac12(L_k^\dagger L_k)^{\T}\otimes\mathbb{1}\big]$. Its eigenvalues are Eq.~\eqref{eq:HSspectrum}; $\hat\Lop_D$ is Hermitian and $\hat\Lop_H$ anti-Hermitian as $4\times4$ matrices, which is the Hilbert-Schmidt statement used in Sec.~\ref{sec:dephased}; the null space is spanned by the vectorized identity; and at $\gamma_\phi=4\Delta$ the Jordan form of $\hat\Lop$ contains the block $\left(\begin{smallmatrix}-2\Delta&1\\0&-2\Delta\end{smallmatrix}\right)$. The Liouvillian gap evaluated at $\gamma_\phi/\Delta=1,2,3,4,5,8$ is $0.5,1,1.5,2,1,0.536$ in units of $\Delta$.

\emph{Odd elastic medium.} Inserting $\bu=(u_L,u_T)e^{\ii kx}$ into $\nabla\cdot\sigma$ gives $-D(\bk)\bu$ with Eq.~\eqref{eq:oddD}; repeating the calculation for $\bk$ at angle $\vartheta$ and rotating to the longitudinal-transverse frame returns the same matrix, confirming isotropy. The discriminant of $D/k^2$ vanishes at $|K^o|=B/2$. With $A=-D(\bk)/\zeta$ and $\Xi=(\kB T/\zeta)\mathbb{1}$, the Lyapunov equation gives
\begin{equation}
\Sigma=\frac{\kB T}{k^2\,\mathcal{N}}\begin{pmatrix}a_L&-BK^o\\-BK^o&a_T\end{pmatrix},
\end{equation}
with $a_L=2(K^o)^2+\mu(B+2\mu)$, $a_T=B^2+3B\mu+2[(K^o)^2+\mu^2]$, and $\mathcal{N}=(B+2\mu)[(K^o)^2+\mu(B+\mu)]$; moreover $A\Sigma-\Sigma A^{\T}=\frac{4\kB TK^o}{\zeta(B+2\mu)}\left(\begin{smallmatrix}0&-1\\1&0\end{smallmatrix}\right)$, nonzero exactly when $K^o\neq0$. Proposition~\ref{prop:GDB}(iii) then yields Eq.~\eqref{eq:oddsigma}.

\section{Verification}
\label{app:verification}

All closed-form results in Eqs.~\eqref{eq:chern}, \eqref{eq:HNspectrum}--\eqref{eq:gcothg}, \eqref{eq:langevin}--\eqref{eq:HSspectrum}, \eqref{eq:oddD}, and \eqref{eq:oddsigma}, together with the incompressible odd-viscosity identity and Corollary~\ref{cor:skin}, were checked symbolically or numerically in Wolfram Language. The script also checks the interpretive claims through explicit counterexamples and conditions: the normality and entropy production of the $N=7$ ring, the real spectrum and nonzero affinity of Example~\ref{ex:3state}, generalized detailed balance for the Langevin oscillator and for coupled resonators at a common temperature (and its failure with all variables taken even), the Hilbert-Schmidt structure and Liouvillian gap of the dephased dimer, the non-quantized integrated curvature at $\nu_o=0$, and the failure of GDB for the odd elastic mode. The script is provided as Supplemental Material, together with an independent Python implementation of the same checks (\texttt{holonomy\_checks.py}, using NumPy, SciPy and SymPy), which reports a pass or fail result for each of its 47 tests and runs in a few seconds. Three tests were added in this version: the open-chain spectrum of Sec.~\ref{sec:HN} with its stationary eigenvalue, the detailed balance forced by a spectrum $\{0,-a,\dots,-a\}$, and the stability range $|k_c|<m\omega_0^2$ of the coupled resonators, without which the Gibbs covariance is not positive definite.


\end{document}